\documentclass[aps,twocolumn,showpacs,preprintnumbers,amsmath,amssymb]{revtex4-1}
\usepackage{latexsym}
\usepackage{graphics}
\usepackage{graphicx}
\usepackage[T1]{fontenc}

\newcommand{\ket}[1]{\mbox{$ | #1 \rangle $}}

\usepackage{hyperref}
\hypersetup{%
  colorlinks,
  plainpages=false,
  pdfpagelabels,
  breaklinks,
  pdfview=Fit,
  bookmarksopen,
  bookmarksnumbered,
  linkcolor=black,
  anchorcolor=black,
  citecolor=red,
  filecolor=black,
  urlcolor=blue
  }%

\begin{document}


\title{On the genesis and evolution of Integrated Quantum Optics}

\author{S\'ebastien Tanzilli}\email{e-mail: sebastien.tanzilli@unice.fr}
\author{Anthony Martin}
\author{Florian Kaiser}
\author{Marc P. De Micheli}
\author{Olivier Alibart}
\author{Daniel B. Ostrowsky}

\affiliation{Laboratoire de Physique de la Mati\`ere Condens\'ee, UMR 6622 du CNRS et de l'Universit\'e de Nice - Sophia Antipolis, Parc Valrose, 06108 Nice Cedex 2, France.}

%

\begin{abstract}
\vspace{1cm}
\begin{tabular}{c c}
 \begin{minipage}{0.5\columnwidth}
 Applications of Integrated Optics to quantum sources, detectors, interfaces, memories and linear optical quantum computing are described in this review. By their inherent compactness, efficiencies, and interconnectability, many of the demonstrated individual devices can clearly serve as building blocks for more complex quantum systems, that could also profit from the incorporation of other guided wave technologies.
\vspace{1cm}

\begin{footnotesize} PPLN waveguide under red laser illumination. A periodically poled lithium niobate (PPLN) waveguide is pumped by a red laser for highly-efficient entangled photon-pair generation at a telecom wavelength by spontaneous parametric down-conversion. Entangled photon-pairs at telecom wavelengths are widely employed for long-distance quantum communication applications. \copyright\,S. Tanzilli.\end{footnotesize}
\end{minipage}&
\begin{minipage}{0.5\columnwidth}
 \includegraphics[width=0.8\columnwidth]{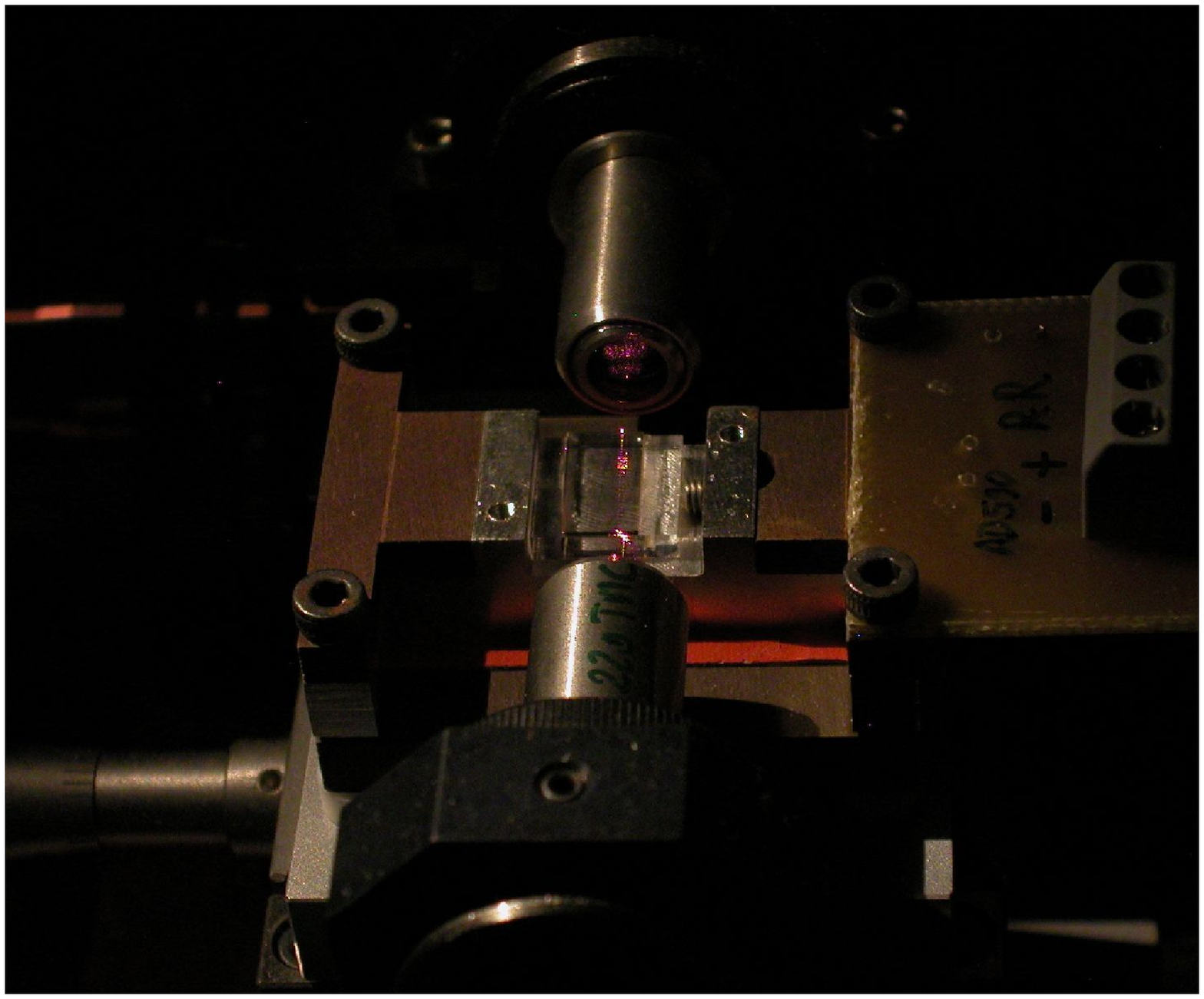}
\end{minipage}
\end{tabular}
\vspace{1cm}
\end{abstract}

 \published{: Laser \& Photonics Reviews DOI: 10.1002/lpor.201100010}
\maketitle
\tableofcontents

\section{Introduction and motivation}

Over the past decade, guided wave optics, in general, and Integrated Optics (IO), in particular, have been writing a new chapter in the ongoing interplay between ``pure'' and ``applied'' physics. In the 1990's new applications and techniques in the field of quantum optics were proposed and developed, based on the exploitation of fundamental quantum properties such as uncertainty and multiparticle superposition (entanglement). It quickly became apparent that the extraordinary technological developments generated by the explosion of fiber optic telecommunications could provide a powerful enabling technology for this emerging field. The object of this paper is to review how and why one of these technologies, \textit{i.e.}, IO, is fulfilling this role.

The new quantum developments usually fall into one of two general fields: quantum computing and/or quantum communications. The quantum computing field has been marked by an enormous number of theoretical papers~\cite{Deutsch85,DiVincenzo95,Nielsen00}. Soon after the seminal paper of Knill-Laflamme-Milburn showing that  quantum computing was feasible using linear optical circuits~\cite{Knill01}, many experimental proofs-of-principle of two-qubit gates~\cite{O'Brien03,Pittman03,Takeuchi10}, were quickly followed by demonstrations based on IO technologies. Beginning with demonstrations of very elementary functions~\cite{Politi08,Marshall09,Smith09,Sansoni_EntangChip_2010}, implementation of more complex algorithms~\cite{AlbertoPoliti09042009,AlbertoPeruzzo09172010,Laing10,Matthews09,Matthews10} soon followed. Quantum communication has been marked by extensive experimental development demonstrating, for example and without being exhaustive, quantum key distribution (QKD)~\cite{BB84,Ekert_Crypto_1991,gisin_QKD_2002,QKD_commercial}, quantum teleportation~\cite{Bennett_Qtele_1993,Bouwmeester_Qtele_1997,Boschi_Qtele_1998,Pan_Swap_1998,kim_Qtele_2001,Marcikic_Qtele_2003,Ursin_Tele_Danube_2004,Landry_Qtele_PlainPalais_2007}, quantum relays based on entanglement swapping~\cite{DeRied_QRelay_2004,Collins_QRelays_2005,DeRiedmatten_swapping_2005,Halder_Ent_Indep_2007,Fulconis_TwoPhotInterf_2007,Kaltenbaek_Inter_Indep_2009,Takesue_Ent_swap_2009,Aboussouan_dipps_2010}, and quantum repeaters~\cite{briegel_quantum_1998,duan_long-distance_2001,simon_quantum_2007}. The development of quantum memories is a subject that impacts both fields~\cite{Shapiro_LongDistTele_2001,Julsgard_QMem_2004,Langer_LongLivedQM_2005,Chaneliere_RemoteAtomEnt_06,staudt_interference_2007,staudt_fidelity_2007,Walmsley_BroadMapping_2007,de_riedmatten_ss_2008,Pan_QMRb_2008,Pan_QMRb_2009,lvovsky_optical_2009,tittel_PEQMSS_2009,simon_quantum_2010,Saglamyurek_BroadbandWQM_2011,Clausen_QSPECrystal_2011}. While all these subjects have used IO technologies, and will be discussed in turn, we begin our paper with a discussion of the communication aspects which have, we believe, been most strongly impacted by IO technology.

The use of guided wave optics as an enabling technology for quantum optics was dramatically demonstrated in the seminal experiment of Nicolas Gisin's Geneva based group in 1998~\cite{tittel_experimental_1998}. In this experiment, fibers in the Swiss Telecom network were used to perform a Bell inequality experiment with the detectors placed in two villages separated by over 10\,km, as shown in Sec.~\ref{Sec_Initial_Stim}. This experiment clearly showed the benefit of the use of optical fibers for transmission of quantum states and strongly suggested that other guided wave optical components, based on IO technologies, could provide a powerful technological base for such experiments~\cite{miya_silica-based_2000,ostrowsky_introduction_2001,honjo_differential-phase-shift_2004}. Over the ensuing years, this proved to be the case, as we shall show in this paper.
The majority of devices we shall discuss have taken advantage of guided wave optics ability to enormousy enhance the efficiency of nonlinear interactions. This is because it permits maintaining high optical power densities over distances far exceeding that permitted by the diffraction limit; some centimeters for IO, many kilometers for fiber optics. In both cases, waveguiding technology revolutionized the field of nonlinear optics. For passive devices IO has provided a possibility of assembling many components on a single chip, enormously simplifying the realization and use of complex circuits.

But one cannot forget a problem which has haunted IO since its birth: that of coupling the IO devices to beams propagating in fibers or free space. This coupling problem has to be taken into account when designing experiments, and, will have to be solved for IO technology to attain its full potential in quantum optics experiments. It is essentially the coupling problem which is responsible for the fact that in some experiments, better results have been obtained using ordinary bulk optics. While these are seminal experiments, it is only through a technology, such as IO, that one can progress towards practical, standardized, low cost, interconnectable, and reconfigurable elements. In this sense, IO should play the role for quantum optics and communication as that played by integrated circuits in electronics. This is why we will be concentrating on IO technology in this paper.

We will now begin by briefly presenting one of the main inspiration for modern quantum optics and communication: The Bell experiment. We will then go on to discuss in more detail, the seminal experiment that served to stimulate the introduction of IO technology into quantum optical experiments. We will then present IO sources, and IO applications to up-conversion detectors, quantum interfaces, quantum relays, components for quantum memories, and hence quantum repeaters, the use of IO for linear quantum computing and finish with a discussion of the perspectives and the conclusion. This review is not intended to be encyclopaedic. What we do intend to do is explain the advantages and disadvantages of integrated optical technology as applied to various tasks, while citing work of historical interest as well as the current state of the art.

\subsection{A fundamental stimulus: Bell experiments}
\label{Sec_fundamental_Stim}

\begin{figure}
\includegraphics[width=1\columnwidth]{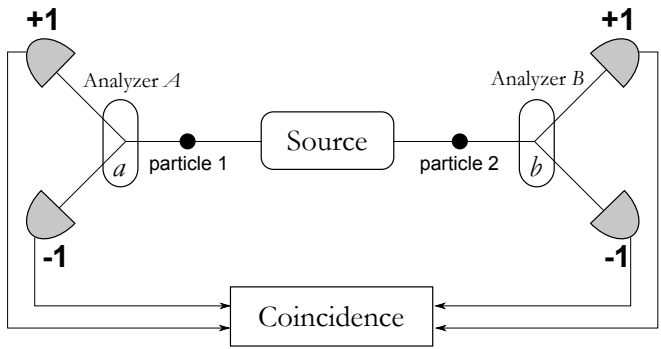}
\caption{General schematic of an EPR experiment where $a$ and $b$ represent the independent settings of analyzers $A$ and $B$, respectively. Note that the results are always $+1$ or $-1$.}\label{Fig_Bell_simple}
\end{figure}

One of the essential inspirations for the recent developments in quantum optics was the demonstration of the violation of the Bell inequalities in experiments involving entangled states of photons. An entangled state is one in which, for example, a two-particle system is in a pure state, leading to quantum correlations, while each separate particle is in an indefinite state. Such states were at the heart of the ``Gedanken Experiment'' proposed by Einstein, Podolsky and Rosen (EPR) in 1935~\cite{EPR_1935} intended to prove the incompleteness of quantum mechanics, since it does not allow the prediction of events by assigning actual values for the properties of the individual particles. They considered this to be a prerequisite of any real situation. However, entanglement leads to the prediction of correlations which violate the Bell inequalities~\cite{Bell_1964} imposed by any local theory and one seems forced to abandon EPR's (and most people's) idea of reality and locality. By locality we mean that events occurring at different points in space cannot influence or affect each other if they have a space-like separation, \textit{i.e.}, the events occur separated by time intervals longer than the time it takes for light to propagate from one point to the other. For the sake of clarity, we now give a brief presentation of a EPR-Bell situation.
Different pairs of non-commuting observables have been used in experiments demonstrating violations of the Bell inequalities. These include polarization, position and momentum, as well as energy and time. An example of such an experiment based on polarization is schematized in Fig.\ref{Fig_Bell_simple}.

In such an experiment an entangled photon pair is created at the source and the individual photons propagate to two separate analyzers, $A$ and $B$, each having two possible outputs and whose settings are defined by the variables $a$ and $b$. The settings could be the angle of a polarization analyzer, the phase difference between the arms of an interferometer, \textit{etc.}, depending on the conception of the experiment. We define the results obtained with analyzer settings $(a,b)$ as $A_a$, $B_b$, and with analyzer settings $(a',b')$ as $A_{a'}$, $B_{b'}$. Values of $+1$ or $-1$ are assigned to events in which photons are detected in the corresponding analyzer outputs and the various coincidences are counted. If we then define the quantity:
\begin{equation}
 M=\left|A_a\left(B_b\pm B_{b'}\right)\right| + \left|A_{a'}\left(B_b\mp B_{b'}\right)\right| \label{eq_mBell}
\end{equation}
It is clear that we have $M \leq 2$  since one of the terms in the interior brackets is equal to zero and the other to $+2$~\cite{Eberhard_78}. Hence the observable expectation value of M, $\langle M \rangle$, is also within this bound. This is the Bell, Clauser, Horne, Shimony and Holt~\cite{BCHSH_1969} form of the Bell inequality. Any local theory ($A$ depends only on $a$, $B$ depends only on $b$), even with an arbitrary distribution of additional unknown, or ``hidden'' variables, must yield results within this bound. However, over the period spanning roughly 1970 to the 1980's, the existence of quantum correlations which violate these inequalities were demonstrated in several experiments with the results becoming ever more precise and convincing~\cite{freedman_experimental_1972,aspect_experimental_1981,Aspect_Bell_time_1982,Aspect_Bell_new_1982}. With a proper choice of analyzer angles, $\langle M \rangle$ can attain the value of $2\sqrt{2}$, which is in perfect agreement with the quantum mechanical prediction when maximally entangled states are considered.

\subsection{The initial stimulus for guided-wave quantum optics: a Bell experiment using existing fiber optic links}
\label{Sec_Initial_Stim}

\begin{figure}
\centering%
\includegraphics[width=\columnwidth]{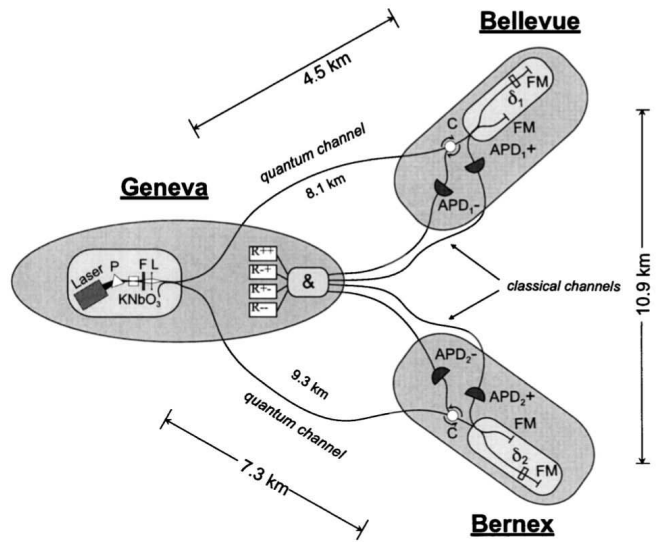}
\caption{Schematic of the Swiss experiment. Figure extracted from Ref.~\cite{tittel_experimental_1998}.\label{Fig_10kms}}
\end{figure}

In 1989, Franson proposed a Bell experiment based on energy conservation that is well-adapted to a fiber optic configuration~\cite{Franson_Bell_1989}. Pioneering demonstrations of this were carried out by a group at Malvern using up to 4\,km of optical fiber coils but with the entire experiment on a single table~\cite{Tapster_Bell_4km_1994}. It was, however, the Geneva group which first demonstrated the ``magical'' non-locality of quantum mechanics with analyzers separated by a distance on the order of 10\,km, as depicted in Fig.\ref{Fig_10kms}~\cite{tittel_experimental_1998}.

In this experiment the laser was a single longitudinal mode diode emitting 8\,mW continuously at 655\,nm. This beam passed through a bulk potassium niobate ($KNbO_3$) crystal which produces parametrically downconverted signal-idler pairs at 1310\,nm. These pairs are \textit{energy-time entangled} since the pair emission time is unknown within the coherence time of the pump laser, and since there is an uncertainty about the individual photon energies although their sum is well known. But once one photon is detected and its energy determined, the energy of the second is also determined. The beams then pass through a modal and spectral filter to eliminate the pump light and limit the signal-idler bandwidth. The pairs are then coupled into a 3\,dB fiber coupler in which half the pairs are split and exit the coupler in separate output branches. Each of these outputs is coupled into standard telecommunication fibers of the Swiss PTT, going to two villages separated by 10.9\,km. The analyzers in this case were optical fiber unbalanced, reflective, Michelson Interferometers with phase modulators in their long arms, and with only one output fiber, connected to an avalanche photodiode (APD). These single photon detectors then trigger laser diodes which emit classical 1\,ns light pulses which are transmitted back to the central station where coincidences are measured.
The coincidences can arise when both of the photons at the distant stations pass through the short arms of the interferometers, or, the long arms of the interferometers. Long-short false coincidences are eliminated by choosing a differential length (20\,cm) between the long and short arms to be larger than the discrimination time of the coincidence counter. Since the coincidences arise from a sum of indistinguishable paths (long-long and short-short), quantum mechanics tells us to sum the probability amplitudes of these events and then take the modulus squared of this amplitude as the probability of the coincident event. 
An analysis of this situation shows that the normalized coincidence probability is given by:
\begin{equation}
 P = \frac{1}{4} \left( 1 + V e^{-\left( \lambda \frac{\phi_1- \phi_2}{2\pi L_c} \right)^2} \cos(\phi_1 +\phi_2) \right)
\end{equation}
where V is an experimental visibility factor and $L_c$ is the individual photon coherence length~\cite{Rarity_Bell_90}. In this experiment the individual photon coherence length was around 10 nm and the long-short arm length distance on the order of 20\,cm. Therefore, when one varies the arm length in either interferometer, changing $\phi_1$ or $\phi_2$, the corresponding single photon counting rate does not change. However, and this is the magical part, the net coincidence count rate varies sinusoidally with an 82\% visibility! It is important to note that a fringe visibility above 71\%, which is the equivalent of M greater than 2 (see Eq.~\ref{eq_mBell}), proves the existence of entanglement and that the quantum correlations persist over 10\,km distances.
In this seminal experiment the source was realized using bulk optics. We will now consider in what ways IO technology could facilitate this and related experiments. We will begin with the most obvious improvement, that of quantum sources, and then discuss other potentially useful components.

\section{Integrated optical sources for quantum optics and communication}
\label{Sec_sources}

In this section, we start by outlining the way in which IO can be used to produce efficient sources for quantum optics experiments. We shall begin with the simplest sources, single photon sources, and then go on to describe photon-pair sources such as that used in the previously discussed experiment, as well as the means of characterizing them.

Single photon sources are necessary for fundamental experiments as well as for certain quantum cryptographic key distribution protocols, mainly based on the BB84 protocol~\cite{BB84,gisin_QKD_2002}. While no perfect single photon state on demand source has been realized to date, a number of approximate versions have been produced. These include weak laser pulses and Heralded Single Photon Sources (HSPS), both of which can be implemented using IO technology, as well as single two-level quantum systems, and quantum dots. 
The essential criteria for comparing such sources are the probability $P_1$ of having a single photon at a specified time, $P_2$, the probability of having two photons rather than one (problematic for quantum cryptography), and $P_0$, the vacuum component, as well as the practical feasibility of the source. To be brief, weak laser pulses suffer from a high $P_2$ and $P_0$, single two-level systems from low collection efficiencies, and quantum dots from usually functioning at cryogenic temperatures. Furthermore, the latter two types of sources usually function at wavelengths that are not compatible with existing telecom fiber networks. However, work in this field is continuing, and while overall collection efficiencies are still relatively low, the progress is promising for future improvements~\cite{Prawer_Diamond_review_2008,dousse_Qdotultrabright_2010,Rivoire_dot_11}.

We now briefly describe the use of IO technology for the realization of a performant weak laser pulse source that was used in association with a Planar Lightwave Circuit (PLC) interferometer for quantum key distribution~\cite{honjo_differential-phase-shift_2004}.
We will follow this with a discussion about approaches based on nonlinear IO devices for the realization of HSPSs and entangled photon-pair sources.

\subsection{Performant QKD experiment based on IO technology}

\begin{figure}
\centering%
\includegraphics[width=\columnwidth]{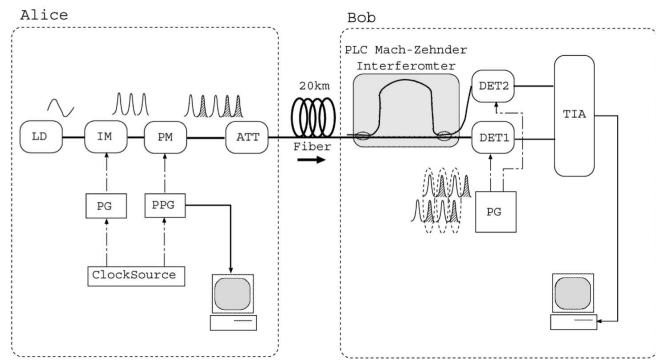}
\caption{\label{Fig_faintQKD}Schematics of a performant QKD system based on faint laser pulses and a PLC circuit. LD represents the laser diode, IM the intensity modulator, PM the phase modulator, ATT the attenuator, PLC the IO Mach-Zehnder circuit, DET 1 and 2 the detectors, and TIA the time interval analyzer. Figure extracted from Ref.~\cite{honjo_differential-phase-shift_2004}.}
\end{figure}

In 2004, a performant quantum key distribution system was demonstrated making extensive use of IO technology~\cite{honjo_differential-phase-shift_2004}. A schematic diagram of this source is shown in Fig.~\ref{Fig_faintQKD}.

In this experiment a CW diode laser (1550\,nm) is modulated by a classical telecom lithium niobate intensity modulator to produce a 1\,Gbit/s train of 125\,ps pulses. These pulses then pass through a lithium niobate (0, $\pi$) phase modulator driven by a random pulse generator. After attenuation to a 0.1\,photon/pulse level the photons were sent through 20\,km of optical fiber and coupled into a Mach-Zehnder IO interferometer based on Silica-On-Silicon (SOS)~\cite{miya_silica-based_2000,yaffe_polarization-independent_1994,Ostrow_73,Ostrow_96} technology having an arm length delay difference equal to the 1\,ns pulse separation time. The single photons exit the interferometer through one output arm or the other depending on the phase imposed by the phase modulator. With both the emitter and receiver parties knowing the arrival time of the photons, this allows the distribution of a key. The overall performance of this configuration was excellent (3000\,bit/s key creation rate with a 5\% quantum bit error rate) validating the extensive use of IO technology, both active and passive. The system was later improved through another use of IO technology~\cite{honjo_field_2007}, that of upconversion detectors, that we will describe in Sec.~\ref{Sec_Up_unitary}.

These upconversion detectors, as well as the sources we will describe in the next sections are all based on nonlinear interactions in IO waveguides which we briefly introduce in the next paragraph.

\subsection{Waveguide based nonlinear optics}
\label{Sec_NLO_waveguide}

Today's HSPS's~\cite{alibart_HSPS_2005}, as well as entangled photon-pair sources we shall discuss later, usually take advantage of photon-pairs produced by Spontaneous Parametric Down Conversion (SPDC) in $\chi^{(2)}$ nonlinear crystals~\cite{burnham_observation_1970}. 
In this interaction, a single pump photon splits into two lower energy photons called the signal and idler photons. For the HSPS, the essential point, as already mentioned, is that the two photons are created simultaneously, or with a known delay, with polarization states imposed by the crystal symmetry and the pump beam polarization. Furthermore, in such a process photon energy and momentum conservation impose:
\begin{equation}
\label{Eq_conservation}
\left\lbrace
\begin{array}{l c l}
\omega_s + \omega_i &=& \omega_p,\\
\overrightarrow{k_s} + \overrightarrow{k_i} &=& \overrightarrow{k_p},
\end{array} \right.
\end{equation}
\noindent
where the indices $\{p,s,i\}$ stand for the pump, signal, and idler photons, respectively.

In the degenerate case, signal and idler photons have equal energies. A continuum of other energy and momentum pairs are possible, determined by the dispersion relation of the material used, but the essential point is that energy and momentum correlations are imposed by these two conservation laws.

\begin{figure}
\centering%
\includegraphics[width=\columnwidth]{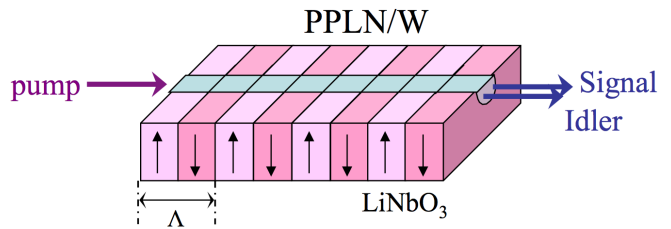}
\caption{\label{Fig_PPLNW}Quasi-phase-matched parametric generation using a PPLN waveguide. $\Lambda$ is the poling period which leads to having an additional grating-type K-vector in the phase-matching equation.}
\end{figure}

The most efficient parametric generators realized to date consist of waveguide structures fabricated in periodically poled lithium niobate (PPLN)~\cite{tanzilli_ppln_2001,tanzilli_ppln_2002} or periodically poled KTP (PPKTP)~\cite{Fiorentino_PP_2007}. In this configuration, shown in Fig.~\ref{Fig_PPLNW}, a periodic reversal of the ferroelectric polarization reverses the sign of the second order nonlinear coefficient to allow Quasi-Phase-Matching (QPM)~\cite{lim_second-harmonic_1989,webjorn_fabrication_1989}.

With this configuration the phase-matching condition becomes $\overrightarrow{k_p} = \overrightarrow{k_s} + \overrightarrow{k_i} + \frac{2\pi}{\Lambda}\cdot\overrightarrow{u}$, where $\Lambda$ is the poling period, and $\overrightarrow{u}$ a unitary vector along the propagation axis. By an appropriate choice of $\Lambda$, one can quasi-phase-match practically any desired nonlinear interaction within the transparency range of the crystal. Moreover, it becomes possible to use the largest nonlinear coefficient of the material. In lithium niobate this is the $d_{33}$ coefficient, which is approximately six times larger than the $d_{31}$ coefficient which must be used for birefringent phase matching in bulk devices, or, when we wish to directly generate pairs with different polarizations.

Taking into account the reduction of the QPM conversion efficiency, compared to birefringent phase matching, of a factor of $\left( \frac{2}{\pi}\right)^2$ we obtain an overall improvement of a factor of 20 in the conversion efficiency. A particularity of configurations using the $d_{33}$ coefficient is that both the signal and idler have the same polarization. If two polarizations are necessary, it is still possible to realize waveguide generators (by titanium indiffusion, for example) using the lower $d_{31}$ or $d_{24}$ coefficients, which will still have higher generation efficiencies than bulk devices for reasons we outline in the following paragraph. 

As previously mentioned, the use of a waveguiding structure permits confinement of the pump beam over the entire interaction length, rather than simply near the focal point of a lens, leading to an efficiency improvement of one to four orders of magnitude when using pumps in the one micron region and sample lengths on the order of centimeters. 
Taking into account the fact that the pump to guide coupling efficiency could be on the order of 25\% (a rather pessimistic value) we still obtain a three order of magnitude improvement over the bulk, birefringently phase matched configuration when using the $d_{33}$ coefficient and a one order of magnitude improvement using $d_{31}$. In addition to the increased generation efficiency, such sources provide a single transverse mode beam that eliminates the problem of post-generation modal filtering. Furthermore, the relatively narrow spectral widths, typically on the order of 40\,nm, compared to the 90\,nm width in the Swiss experiment, reduces by a factor 2 the need for optical bandwidth limiting of the signal and idler beams. The overall improvement in the raw efficiency can be of 4 orders of magnitude, as was demonstrated in experiments performed in the following years~\cite{tanzilli_ppln_2001,sanaka_new_2001}. 

Most of these first IO source experiments were carried out using Proton Exchange (PE) to fabricate guides on Z-cut lithium niobate typically realized by substrate immersion in a benzoic acid bath heated to around 300$^\circ$C during 72\,h~\cite{Chanvillard_SPE_2000}. This permitted achieving the expected 3 to 4 orders of magnitude of improvement, when compared to bulk sources, resulting in a conversion efficiency (probability of pair creation per pump photon) of about $10^{-6}$ at telecom wavelengths (1310 or 1550\,nm). Similar results were obtained near 800\,nm using PPKTP~\cite{banaszek_generation_2001}. Such rubidium, caesium, or thallium ion-exchanged IO devices were initiated in the late 80's by J.D. Bierlein and collaborators, and good examples can be found in Refs.~\cite{Bierlein_KTP_87,Bierlein_KTP_92,Bierlein_KTP_94} regarding both fabrication processes and characterization techniques.

\subsection{Heralded Single Photon Source}

The next type of source we describe is the HSPS. Regarding the history, the idea of the HSPS, \textit{i.e.}, taking advantage of the detection of one single photon, of the two generated simultaneously, to herald the emission time of the second photon, is not new.
To our knowledge, the first HSPS was demonstrated in 1986~\cite{Grangier_HSPS_1986}, after a set of experiments performed in Orsay~\cite{aspect_experimental_1981,Aspect_Bell_time_1982} aimed at demonstrating the violation of the so-called Bell inequalities~\cite{Bell_1964,BCHSH_1969,EPR_1935}. This HSPS was made using polarization entangled photons emitted by a double radiative cascade in calcium atoms. Using this source, the Orsay group demonstrated single photon interference patterns showing almost perfect visibilities~\cite{Grangier_HSPS_1986}.

\begin{figure}
\centering%
\includegraphics[width=\columnwidth]{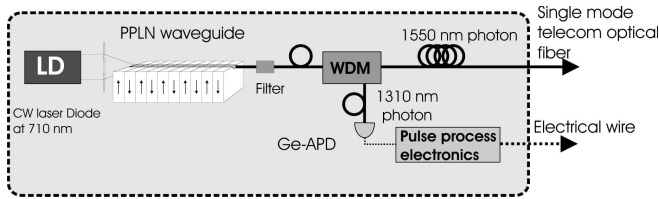}
\caption{\label{Fig_HSPS}A Heralded Photon Source (HSPS) built around a PPLN waveguide generating single photons at 1550\,nm upon the detection of trigger photons at 1310\,nm using a germanium avalanche photodiode (Ge-APD). Figure extracted from Ref.~\cite{alibart_HSPS_2005}.}
\end{figure}

The HSPS we now describe is based on the use of photon-pairs generated in a PPLN waveguide realized by PE. A schematic of the overall experiment is shown in Fig.~\ref{Fig_HSPS}.

In the HSPS, a 710\,nm laser, operating continuously, produces entangled pairs of photons at 1310 and 1550\,nm respectively. These are coupled into a standard single-mode optical fiber and then passed through a telecom fiber-optic Wavelength Division Multiplexer (WDM) which separates the photons. The 1310\,nm photon is connected to a Ge-APD and upon detection a pulse is generated that announces the arrival (after delay by one hundred meters of fiber) of the 1550\,nm photon in the single-mode telecom output fiber. This is the desired single photon that can be used for quantum cryptography, or other quantum manipulations, implementing transmission over existing telecom networks. By measuring the counts at the output of the fiber when the trigger photon has been detected we measured the conditional $P_1$ to be 37\%. This is mainly due to the low (bulk optics) coupling efficiency between the guide and the fiber (58\%), and dark counts from the Ge-APD that impacts the parameter $P_0$. Both of these limitations can be overcome, to some extent, by the use of existing technologies. The collection efficiency could be seriously improved by correctly pigtailing the output fiber to the waveguide, and the dark counts could be greatly reduced by generating a photon-pair at 810 and 1550\,nm, respectively. This would enable using a silicon APD to detect the trigger photons, greatly reducing the dark counts, and increasing the detection efficiency, which has already been demonstrated with bulk optics~\cite{Fasel_HSPS_2004}.

Inserting a beam-splitter after the fiber output and counting coincidences allows measuring $P_2$ which was on the order of $5\times10^{-3}$ in our experiments, which for equivalent $P_1$, is four times better than with weak laser pulses~\cite{alibart_HSPS_2005}.
Nevertheless, the heralded photons produced by the above-mentioned devices are not in a true single photon state. Such a state would be a Fock state with only one photon, in the absence of any vacuum state or multi-photon state contributions. While, to the best of our knowledge, no such generator has been realized to date, there have been several suggestions made based on cross-phase modulation~\cite{konrad_production_2006} or type-II downconversion in a KTP waveguide without periodic poling~\cite{levine_heralded_2010}. Both propositions present interesting possibilities for future developments.

The only essential characteristic of photon-pairs necessary for the functioning of HSPS type sources is the emission of the two photons, simultaneously, or with a programmable delay between them as in the Duan-Lukin-Cirac-Zoller (DLCZ) protocol~\cite{duan_long-distance_2001,Sangouard_DLCZRMP_2011}. We will now go on to discuss the cases where \textit{entanglement} is an essential requirement.

\subsection{Entanglement sources, photon-pair production and characterization}

The demonstration of the use of IO technologies to remarkably increase photon-pair production efficiencies was rapidly followed by the demonstration of pair entanglement (quantum interference), at both telecom and visible wavelengths. When comparing in detail the various devices we will discuss, one has to keep in mind their intended applications. Nevertheless, the recent trend has been in the direction of maximizing spectral brightness, to permit very narrow filtering, while maintaining a relatively high efficiency and acceptable counting rates in experiments. IO sources are so efficient that it is necessary to avoid creating multiple pairs. Already published review papers of the field can be found in Refs.~\cite{weihs_photonic_2001,gisin_quantum_2007,Beveratos_Tanzilli_EBQC_2008}.

\begin{figure}
\centering%
\includegraphics[width=\columnwidth]{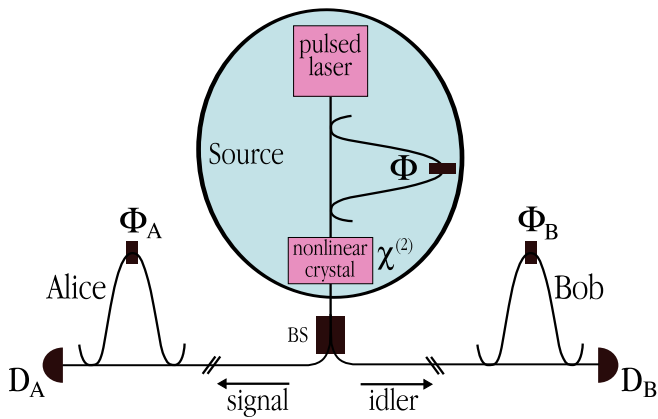}
\caption{\label{Fig_TB}Simplified setup for generating time-bin entangled photon-pairs. A preparation interferometer placed in the path of the pump pulses permits creating coherent superpositions of time-tagged states, of the form $\ket{\Psi_{pump}}=\frac{1}{\sqrt{2}}(\ket{short}+e^{i\phi}\ket{long})$. In the following, a photon taking the short path is identified as a $\ket{0}$ and a photon taking the long path is identified as a $\ket{1}$. Then, when such a state superposition goes through a nonlinear crystal, photon-pairs can be emitted either from the early or from the late pump pulse possibilities, subsequently leading to an entangled state of the form $\ket{\Psi_{pair}}=\frac{1}{\sqrt{2}}(\ket{0}_s\ket{0}_i+e^{i\phi}\ket{1}_s\ket{1}_i)$. In this scheme, two other interferometers, having varying phases $\phi_A$ and $\phi_B$, respectively, are necessary for revealing entanglement, as is the case in the Franson configuration~\cite{Franson_Bell_1989} (see Sec.~\ref{Sec_Initial_Stim}). Figure extracted from Ref.~\cite{Beveratos_Tanzilli_EBQC_2008}.}
\end{figure}

IO based entanglement sources naturally followed earlier reported experiments aimed at generating energy-time entanglement based on CW pump lasers (see Sec.~\ref{Sec_Initial_Stim} and Ref.~\cite{tittel_experimental_1998}), or a powerful variant called time-bin entanglement. As opposed to energy-time, time-bin entanglement enables having clocked experimental realizations since it permits the use of pulsed pump lasers through the implementation of a preparation interferometer to create time-tagged coherent superpositions of states, as depicted in Fig.~\ref{Fig_TB} (see caption for details) and discussed in the seminal paper published in 1999 by the Geneva group~\cite{Brendel_TB_1999}.

In the following years, many other entangled photon pair sources were demonstrated. Among them were:
\begin{itemize}
\item In 2001, two new sources based on type-0 (the three fields have identical polarizations) PE:PPLN waveguides, one from a Tokyo group~\cite{sanaka_new_2001}, the other from a Geneva/Nice collaboration~\cite{tanzilli_ppln_2001}, were demonstrated. The source from Tokyo dealt with photons at 854\,nm and led to a two-photon interference raw visibility of 80\% when testing energy-time entanglement in a Franson configuration~\cite{tittel_experimental_1998,Franson_Bell_1989} (see Sec.~\ref{Sec_Initial_Stim} for more details on Franson-type experiments).\\
Conversely, the Geneva/Nice source dealt with photons in the telecom range (1310\,nm)~\cite{tanzilli_ppln_2001} and led, in 2002, to demonstrations of energy-time and time-bin entanglement showing 97\% and 84\% net visibilities, respectively, when a Franson configuration was used to test the entangled states~\cite{tanzilli_ppln_2002}. The latter source was then used for demonstrating the distribution and preservation of non-maximally time-bin entangled states over 11\,km of optical fiber.
\item More recently, a slightly modified version of this source, emitting paired photons at 1310 and 1550\,nm wavelengths, led to a 98\% net visibility two-photon interference for maximally energy-time entangled states. This source was the initial key element for the demonstration of the first entanglement quantum interface which we will discuss in Sec.~\ref{Sec_Up_unitary}~\cite{Tanz_Interface_2005}. With these new IO sources, a common feature demonstrated was a much higher raw pair production efficiency, \textit{i.e.}, an improvement of more than 4 orders of magnitude when compared with sources based on bulk nonlinear crystals~\cite{tanzilli_ppln_2001}. This opened the way to more complex experiments, \textit{e.g.}, in which more than two photon-pairs are to be produced at the same time, or when very narrow photon-pair filtering is necessary.
\item Following this idea, a new generation energy-time entanglement source was subsequently demonstrated by the Geneva group in 2007, aimed at entangling independent photons by time detection in a quantum relay configuration (see Sec.~\ref{Sec_QRelay} for more details on quantum relay operations). Such an experiment required extremely long coherence time photons obtained by dramatically filtering ($\sim$10\,pm) light emitted by PPLN waveguide based sources. This permitted the demonstration of two-photon quantum interference, the so-called Hong-Ou-Mandel (HOM) dip~\cite{HOM_dip_1987}, using photons from two autonomous sources, pumped by different rubidium-stabilized CW lasers~\cite{halder_high_2008}.
\item Several years after the first IO based energy-time entanglement sources appeared, the first IO based polarization entanglement sources were announced. That was an important step since the seminal papers, dealing with photon-pairs emitted via double atomic radiative cascades and demonstrating violation of the Bell inequalities were based on polarization entangled states~\cite{freedman_experimental_1972,aspect_experimental_1981}. Note also that many polarization entanglement sources based on bulk nonlinear crystals were reported in the literature. They proved that nonlinear optics could be a very efficient, reliable, and compact means to generate entanglement. Good examples are given in Refs.~\cite{Kwiat_HIntesity_1995,kwiat_ultrabright_1999,fedrizzi_wavelengthtunable_2007,Fiorentino_PP_2007}.
\item In 2003, the first demonstration of an IO based polarization entanglement source was made using two PE:PPLN waveguides mounted in a Mach-Zehnder interferometer configuration. This allows superimposing, using suitable half wave-plates and polarization beam-splitters, pairs of orthogonal states of polarization~\cite{yoshizawa_generation_2003}. This way, a pair of photons coming out of such a device are either in the $\ket{H,H}$ state or in the $\ket{V,V}$ state, with equal probabilty amplitudes, therefore producing the Bell state $\ket{\Phi^\pm} = \frac{1}{\sqrt{2}} \left[ \ket{H,H} \pm \ket{V,V}\right]$. The paired photons were generated at 1550\,nm with a bandwidth of 30\,nm. By applying a phase shift in one arm of the interferometer, a net visibility of 94\%  was obtained.
\item In 2005, a means of converting time-bin entanglement to polarization entanglement was demonstrated, but with a limited net visibility of 82\%~\cite{takesue_generation_2005}. A pair of horizontally polarized photons, that have been time-bin entangled using an IO amplitude modulator in conjunction with a CW source, enters the structure depicted in Fig.~\ref{Fig_TB_to_polar}. In this case, post-selection permits extracting the desired polarization entangled state.
\begin{figure}
\centering%
\includegraphics[width=\columnwidth]{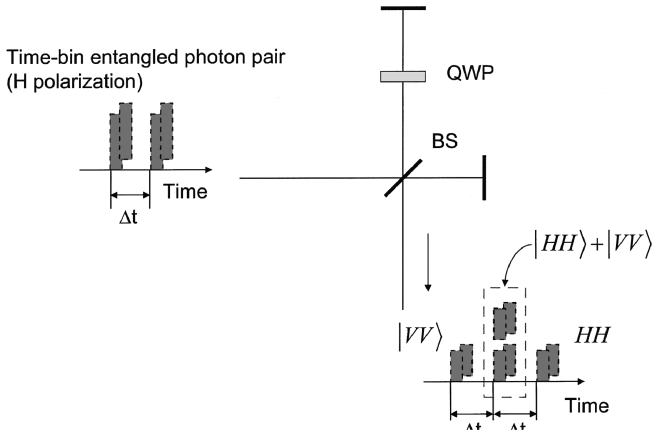}
\caption{\label{Fig_TB_to_polar}Schematic diagram of the time-bin to polarization entanglement conversion experiment. The conversion is based on a Michelson-type arrangement, taking advantage of a beam-splitter (BS) and a quarter-wave plate (QWP). Figure extracted from Ref.~\cite{takesue_generation_2005}.}
\end{figure}
\item In 2007, a polarization entanglement source using two cascaded type-I PPLN waveguides mounted in a Sagnac fiber loop to permit generating 757\,nm pump photons via two-photon upconversion from 1515\,nm, followed by non-degenerate pair generation at 1431 and 1611\,nm was realized, again to produce the Bell state $\ket{\Phi^\pm}$~\cite{jiang_generation_2007}. A net visibility of over 90\% was obtained with this device.
\item Again in 2007, the first demonstrations of generating polarization-entangled pairs using type-II phase matching in a Ti indiffused PPLN waveguide was reported and used to produce the Bell state $\ket{\Psi^\pm} = \frac{1}{\sqrt{2}} \left[ \ket{H,V} \pm \ket{V,H}\right]$~\cite{suhara_generation_2007,fujii_bright_2007}. While, as expected, the efficiencies where lower than that obtained with type-0 interactions due to the imposed use of a less efficient nonlinear coefficient, a net visibility of over 90\% in a Bell test, with an overall efficiency of 5.3$\cdot$10$^{-10}$ and a spectral width of 1\,nm, was obtained in Ref.~\cite{suhara_generation_2007}.
\item In 2010, our group, in collaboration with researchers from the University of Paderborn, was able to produce pairs in this configuration that were characterized with both HOM-dip and Bell measurements, and net visibilities of nearly 100\% were obtained in both cases~\cite{martin_integrated_2009,martin_polar_2010}, using a discernability compensation stage. In this realization depicted in Fig.~\ref{Fig_AM_NJP}, the photon's bandwidth was 0.5\,nm for an overall conversion efficiency of 1.1$\cdot$10$^{-9}$.
\begin{figure}
\centering%
\includegraphics[width=\columnwidth]{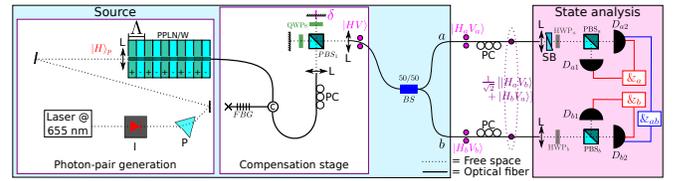}
\caption{\label{Fig_AM_NJP}Experimental setup of the source and analysis system from Ref.~\cite{martin_polar_2010}. The pairs of circles represent pairs of photons with their associated polarization quantum states, after the birefringence compensation system and after the beam-splitter (output of the source). P: prism used to filter out any infrared light coming from the laser; I: optical isolator; L: lenses; PPLN/W: periodically poled lithium niobate waveguide; FBG: tunable fiber Bragg grating filter centered at 1310\,nm with a bandpass of 0.5\,nm; C: fiber optical circulator; PC: polarization controllers; PBS: polarization beam-splitters; QWP: quarter wave-plates; BS: 50/50 coupler; SB: Soleil-Babinet compensator; HWP: half wave-plates; D: single photon detectors; and $\&$: AND-gate that can be placed between any two detectors depending on the performed measurement. Figure extracted from Ref.~\cite{martin_polar_2010}.}
\end{figure}
\item Another experiment based on a PPKTP waveguide, showed performances as high as these~\cite{zhong_high_2009,zhong_highquality_2010}.
All the above mentioned sources employ perfectly degenerate paired photon; otherwise entanglement would be partially or totally lost. They rely on a beam-splitter to separate the twins, leading to a loss of 50\% of the created pairs.
To compensate for this loss issue, a cascade of two QPM gratings was used on a Ti:PPLN waveguide to produce non degenerate paired photons still sharing the $\ket{\Psi^\pm}$ Bell state~\cite{suhara_doubleQPM_2009}. Here, the key point lies in the fact that the two gratings offer two phase-matching conditions and non-degenerate polarization entanglement is found where the two phase-matching curves cross each other. In this case, a net visibility of 70\% was obtained in a Bell test.
Following the same idea, note that a theoretical proposal has been reported in Ref.~\cite{Thyagarajan_polar_2009}, in which polarization entangled photons at non-degenerate wavelengths can be produced using a type-II aperiodically poled lithium niobate waveguide, \textit{i.e.}, using a poling structure that is the convolution of two grating periods. Both the latter schemes greatly simplify the one discussed above involving two cascaded type-I PPLN waveguides, releasing the constraint of having an interferometric setup~\cite{jiang_generation_2007}.
All these demonstrations clearly indicate the promise of such IO devices as sources for quantum communication protocols, in terms of high generation efficiencies but also in terms of QPM engineering.
Finally note that Ref.~\cite{martin_polar_2010} provides the reader with a quite exhaustive review of such type-II sources.
\item In 2007, the first photon-pair source incorporating functional integration appeared. In this device, reverse PE guides were used to realize a device having a two-mode section in the PPLN followed by an adiabatic Y junction serving as an integrated modal beam-splitter~\cite{zhang_correlated_2007}. A schematic of the device is given in Fig.~\ref{Fig_Y_junction}.
\begin{figure}
\centering%
\includegraphics[width=\columnwidth]{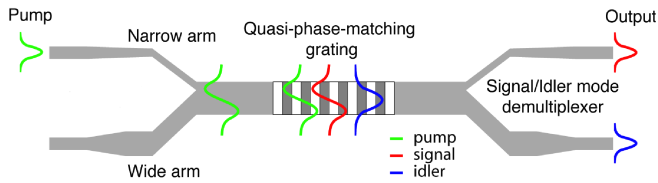}
\caption{\label{Fig_Y_junction}Asymmetric Y junction device for parametric down conversion. Figure extracted from Ref.~\cite{zhang_correlated_2007}.}
\end{figure}
A 780\,nm wavelength pump beam is coupled in the TM$_{00}$ mode through the narrow input arm of the Y-junction and is adiabatically converted into TM$_{10}$ mode of the central waveguide containing the QPM grating. The pump then traverses the QPM grating generating photon-pairs in different spatial modes (signal in TM$_{10}$ and idler in TM$_{00}$, respectively). The idler photon in the TM$_{00}$ mode exits via the wide branch of the Y-junction's output port, while the signal photon in the TM$_{10}$ mode will be coupled into the narrow output branch. Both output modes are then adiabatically transformed to identical TM$_{00}$ modes. Reverse PE results in a somewhat buried and more cylindrically symmetric guide which should improve guide-fiber coupling. The device was operated at 10\,GHz and had a coincidence to accidental pair counting ratio of over 4000, but no measurements of entanglement or efficiency were reported.
\end{itemize}

\subsection{Other source technologies}

In this section, we will briefly cite several interesting related types of sources.

Following a suggestion of Lin and Agarwal~\cite{lin_silicon_2006} some sources have been realized based on a relatively new IO technology, that of silicon nano-wire waveguides~\cite{sharping_generation_2006,harada_generation_2008,Clemmen_Gene_Nano_2010}. Typically, guides are fabricated on a Silicon-On-Insulator (SOI) wafer with a Si upper layer deposited upon a several micron thick layer of SiO$_2$. The guides are several hundred nm thick and on the order of 500\,nm wide with lengths on a cm scale. Such structures push the idea of high confinement to an extreme, since they result in sub-micronic cross-section ($\simeq$0.1\,$\mu$m$^2$) guides leading to extremely high power densities. The pairs are generated by four-wave mixing in the guides and effective nonlinearities as high as $10^5$ (W.km)$^{-1}$have been reported due to the high power density~\cite{harada_generation_2008}. Quantum measurements have led to fringe visibilities on the order of 95\% without correction~\cite{fukuda_ultrasmall_2006}. There remains a problem of spontaneous Raman noise with these devices but they certainly suggest a new direction for both active and passive IO technology. 

Another interesting source, based on an integrated AlGaAs semiconductor ridge microcavity, emitting counter-propagating twin photons in the telecom range has recently been demonstrated. Relying on type-II SPDC and working at room temperature, the latter source produced twin photons showing a natural bandwidth as narrow as 0.3\,nm, which led to a visibility figure of 85\% when measured via a HOM setup~\cite{caillet_counter2phot_2010}.

Note that semiconductor quantum dots surrounded by micropillars have also been widely investigated over the past ten years. Although the related developments are beyond the scope of this paper, the reader can find pertinent examples of this very active field in Refs.~\cite{stevenson_Qdotsource_2006,dousse_Qdotultrabright_2010,claudon_Qdothighly_2010,mohana_Qdotpolarent_2010,patel_Qdottwophoton_2010}.

Another type of single photon emitters have been developed using nitrogen-vacancy centers in diamond. The reader can find interesting and recent examples in Refs.~\cite{Prawer_Diamond_review_2008,Alleaume_NV_2004,Aharonovich_NV_2009,Siyushev_nanodiamond_2009}.

Finally, there have been many sources based on non-linear $\chi^{(3)}$ fibers as well, either relying on Dispersion-Shifted Fiber (DSF) or Photonic Crystal Fiber (PCF) technologies, but only a few emitting at a telecom wavelength~\cite{mcmillan_PCF_2009,Slater_PCF_10,medic_DSF_2010}. These fibers have been employed to realize either HSPS or entanglement sources.
One of the most relevant and recent realizations produces telecom entangled photons via bichromatic pulses pumping a DSF placed in a Sagnac loop aligned to deterministically separate degenerate photon pairs~\cite{medic_DSF_2010}. The authors obtained a state fidelity to the nearest maximally polarization entangled state of 0.997 when subtracting background Raman photons, multiple pair generation events, and accidental coincidences initiated by detector dark counts. In this experiment, note that the Raman effect was globally reduced by cooling the fiber to a temperature of 77\,K (liquid nitrogen). 

\section{Applications of integrated optical sources and guided-wave optics as enabling technologies for quantum relay operations}
\label{Sec_QRelay}

For the realization of quantum networks, interference between photons produced by independent, or autonomous, sources is necessary. From the fundamental side, photon coalescence (or two-photon interference) lies at the heart of various quantum operations, such as quantum logic gates (see Sec.~\ref{Sec_QComp}) and relay functions~\cite{HOM_dip_1987,Rarity_dip_1989,Rarity_dip_1990,LegeroRempe_coalecence_2003,Halder_2times25km_2005}. In particular, this purely quantum effect is seen as a preliminary step towards achieving quantum teleportation~\cite{Bennett_Qtele_1993,Bouwmeester_Qtele_1997,Boschi_Qtele_1998,Pan_Swap_1998,kim_Qtele_2001,Marcikic_Qtele_2003,Ursin_Tele_Danube_2004,Landry_Qtele_PlainPalais_2007} and entanglement swapping~\cite{DeRied_QRelay_2004,DeRiedmatten_swapping_2005,Halder_Ent_Indep_2007,Fulconis_TwoPhotInterf_2007,Kaltenbaek_Inter_Indep_2009,Takesue_Ent_swap_2009,Aboussouan_dipps_2010} that are the fundamental protocols underlying quantum relays~\cite{Collins_QRelays_2005}.

From a more applied side however, as we will discuss in Section~\ref{Sec_Up_det}, the maximum achievable distance for quantum communication links is mainly limited by the inherent dark counts in the employed single photon detectors. To extend the distance, photon amplification, as used in standard telecommunication networks, is not possible since the need here is not to duplicate the number of photons, but rather the qubit they carry along the quantum channel. It is known that copying, or cloning, an unknown qubit with a perfect fidelity is prohibited by quantum physics~\cite{wootters_cloning_1982,scarani_cloning_2005}. Note that the no-cloning theorem, as initially depicted in Ref.~\cite{wootters_cloning_1982}, lies at the heart of QKD in the sense that it prevents an eavesdropper from copying the distributed qubits between the authorized parties~\cite{BB84,gisin_QKD_2002}.

\subsection{Quantum relays based on the entanglement swapping scheme}

\begin{figure}
\centering%
\begin{tabular}{c}
\includegraphics[width=0.95\columnwidth]{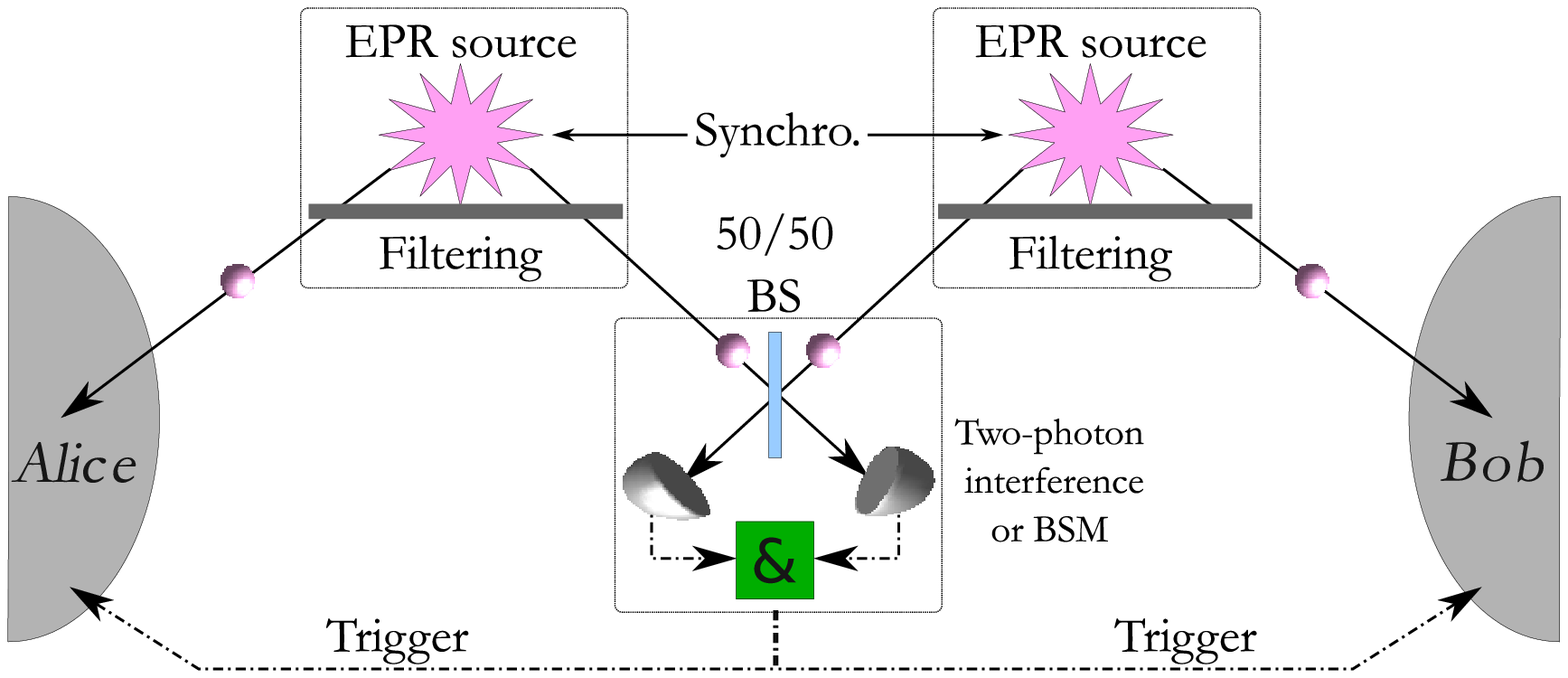}\\
(a)\\
\includegraphics[width=0.95\columnwidth]{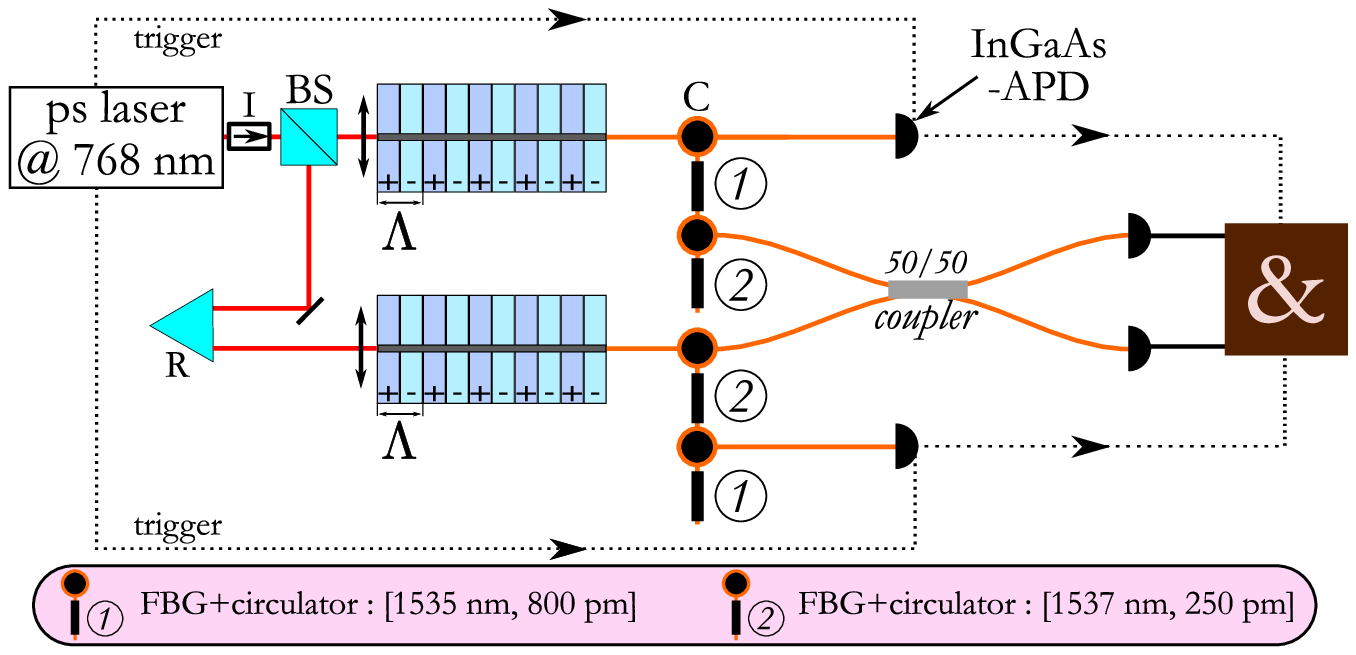}\\
(b)
\end{tabular}
\caption{\label{Fig_Qrelay}
(a) Principle of the quantum relay scheme based on entanglement swapping. This scheme involves two pairs of entangled photons emitted by two synchronized sources and their associated spectral filtering stages. The two inner photons are sent to a 50/50 beam-splitter (BS) where a Bell state measurement (BSM), based on two-photon interference, is performed. Using dedicated electronics, entanglement can be swapped from these photons to the outer ones making Alice and Bob connected by entanglement, as if they had each received one photon from an entangled pair directly. The BSM serves as a trigger for Alice and Bob's detectors so that four-fold coincidences are recorded.
(b) Example of a proof-of-principle quantum relay experiment developed in our group operating at telecom wavelengths for long distance application. The setup is based on a single ps-regime laser and two PE:PPLN waveguide sources. The filtering stages are made using performant telecom components such as fiber Bragg gratings (FBG), fiber-optics circulators (C), and a 50/50 fiber coupler. The photon-pairs are emitted around 1550\,nm and detection is ensured by InGaAs-APDs.
Figure extracted from Ref.~\cite{Aboussouan_dipps_2010}.}
\end{figure}

Physicists soon realized that quantum teleportation and entanglement swapping could provide a means towards extending the maximum achievable distance of quantum communication links~\cite{Collins_QRelays_2005}. To help the reader understand the related key points, Fig.~\ref{Fig_Qrelay} presents the principle of a quantum relay scheme based on an entanglement swapping configuration. Here two autonomous, properly synchronized, entanglement sources, associated photon-pair filtering stages, and a 50/50 beam-splitter are interconnected. One photon from each source, the outer one, is sent to one of the users, \textit{i.e.}, Alice or Bob, while the two inner photons are sent to the beam-splitter. This beam-splitter enables coupling the two latter photons to provide the necessary entangling operation preliminary to the swapping scheme and relay function (see caption for details). 
In principle, the quantum relay function enables reducing the SNR of a quantum link by ``simply'' projecting, upon a joint measurement, two inner photonic qubits that have no common past onto an entangled Bell state. This so-called Bell state measurement (BSM) can subsequently serve as a trigger for the final detectors placed at both ends of the link, therefore improving the SNR and making it possible to increase the maximum achievable distance. This is another example where fundamental principles turn out to be attractive for ``real world'' applications.

The first key point to note in this schematic is the fact that the two entanglement sources have to be properly synchronized so as to avoid any temporal distinguishability between the two inner photons that are subjected to the BSM. Otherwise, the resulting entanglement carried by the two outer photons would be imperfect.
Theoretically, the two inner photons can come from any type of source, provided they are identical at the beam-splitter (BS) (pre-selection) or at the detectors (post-selection).
From the experimental side, we have to compare their coherence time, $\tau_{\mbox{\footnotesize coh}}$, to the time uncertainty, $t_{\mbox{\footnotesize uncert}}$, within which they are created (\textit{i.e.}, the pulse duration of the pump laser(s)) or are detected (detector's timing jitters), which can be written as:
\begin{equation}
\tau_{\mbox{\footnotesize coh}}\geq t_{\mbox{\footnotesize uncert}}.\label{Eq_temp}
\end{equation}
Suitable bandpass filters are therefore employed to achieve optimal interference visibilities. Up to now, this issue has been addressed using different approaches based on pulsed or CW lasers.

Synchronizing two remote sources is far from simple and is highly dependent on the employed operation regime, \textit{i.e.}, whether the pump lasers are CW or delivering femtosecond (fs) pulses or picosecond (ps) pulses. Of course, the CW regime requires no a priori synchronization at all, whereas two fs or ps lasers necessitate, at least, several fs and sub-ps synchronization accuracy, respectively~\cite{Kaltenbaek_Inter_Indep_2006,Kaltenbaek_Inter_Indep_2009,Aboussouan_dipps_2010}.
Without being exhaustive, there have been several reports of entanglement swapping schemes for quantum relay applications based on either bulk crystal sources~\cite{DeRied_QRelay_2004,DeRiedmatten_swapping_2005,Kaltenbaek_Inter_Indep_2009} or guided-wave (crystal or fiber) sources~\cite{Halder_Ent_Indep_2007,Fulconis_TwoPhotInterf_2007,Takesue_Ent_swap_2009,Aboussouan_dipps_2010}.
Basically, bulk crystal generators were associated with the fs regime of operation, whereas guided-wave generators were mainly associated with CW and ps regimes. This is due to the fact that CW and ps regimes require much narrower spectral filtering stages than the fs regime. Orders of magnitude for these filters are several nm, hundreds of pm, and tens of pm, for the fs, ps, and CW regimes, respectively.
These filtering stages represent the second key element since they prevent, in addition to the synchronization procedure, any temporal distinguishability between the inner photons (see Fig.~\ref{Fig_Qrelay}).  A pertinent discussion of these details can be found in Ref.~\cite{Aboussouan_dipps_2010}. As a consequence, it is therefore necessary to employ sources showing much higher photon-pair generation efficiencies to permit reasonable four-fold coincidence rates. In other words, it is the normalized brightness, \textit{i.e.}, the number of pairs generated per second, per mW of average pump power, and per GHz of bandwidth, which matters for the sources. Evidently, one has to take care to avoid multiple pair production which would reduce the quality of the measured entanglement. For more details, see Refs.~\cite{Scarani_2pairs_04,Smirr_2pairs_11}.

In the following, we shall focus on guided-wave realizations, \textit{i.e.}, implying sources based on IO or fiber optical technologies, for which high photon-pair brightnesses were a third key ingredient necessary to obtain coincidence rates as high as possible, both in the CW and ps regimes of operation.

In 2007, the Geneva Group demonstrated a fully independent source quantum relay in the CW regime~\cite{Halder_Ent_Indep_2007}.
Using two CW 780\,nm lasers stabilized against rubidium cells and two PE:PPLN waveguides, energy-time entangled photons were generated around 1560\,nm and sent to fiber Bragg grating (FBG) filters having bandwidths as narrow as 10\,pm for each source. They obtained a two-photon interference pattern showing 84\% net visibility for the inner photons when indistinguishability was tested in the HOM configuration~\cite{HOM_dip_1987}, and subsequently an interference pattern of 63\% visibility when the resulting swapped entanglement was measured on the outer photons using a Bell test in the Franson configuration~\cite{Franson_Bell_1989}. These results are very promising since the two sources are truly autonomous.

In 2007 and 2009, two fiber-optic solutions have been implemented, one relying on microstructured fibers emitting photons in the visible band associated with 300\,pm filters (Bristol University)~\cite{Fulconis_TwoPhotInterf_2007}, and the other on DSFs emitting photons in the telecom range associated with 200\,pm FBG filters by an NTT group~\cite{Takesue_Ent_swap_2009}. It is also worth noting the demonstration of intrinsically narrowband fiber sources allowing heralding one photon in a pure spectral state and, hence, rendering narrowband filtering unnecessary~\cite{PCFHalder09}. In each experiment, two separated fiber generators were pumped using a single ps regime laser. Both experiments showed HOM-dips having raw visibilities of 88, 64 and 76\%, respectively. In addition, the NTT experiment showed an interference pattern of 41\% raw visibility when the resulting swapped entanglement was measured on the outer photons using a Bell test.

\begin{figure}
\centering%
\includegraphics[width=\columnwidth]{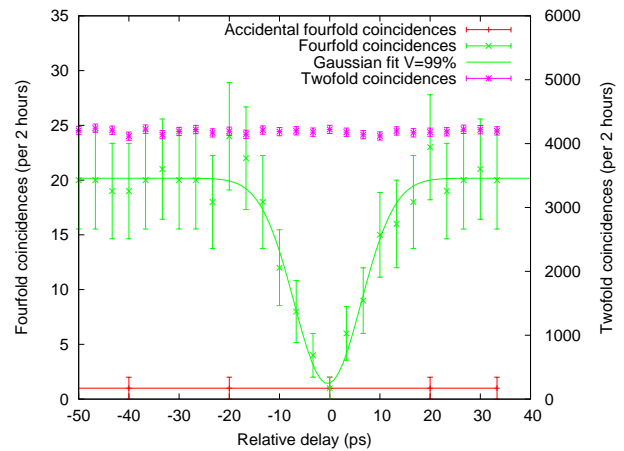}
\caption{\label{Fig_Qrelay_Nice}HOM-dip result for the experiment of Fig.~\ref{Fig_Qrelay}(b), where the fourfold coincidence rate is plotted as a function of the relative delay, $\delta t$, between the interfering photons. We clearly observe a dip for $\delta t = 0$ that reaches the noise level. The Gaussian fit of the interference pattern shows a net visibility of $99\%\pm3\%$. One of the two-fold coincidences, related to one of the sources, after the BS, is used to verify that this figure is constant. Figure extracted from Ref.~\cite{Aboussouan_dipps_2010}.}
\end{figure}

More recently, our group realized an experiment based on two separated PE:PPLN waveguides pumped by a single ps regime laser. Pairs of photons around 1550\,nm were generated in each waveguide and the inner photons (see Fig.~\ref{Fig_Qrelay}(b)) filtered down to 200\,pm thanks to FBG filters. The obtained results were of excellent quality since two-photon interference was observed in a HOM configuration, with a dip having a net (raw) visibility as high as 99\% (93\%), as shown in Fig.~\ref{Fig_Qrelay_Nice}~\cite{Aboussouan_dipps_2010}.

Such proof-of-principle experiments emphasize why IO and guided-wave technologies, implemented both in the ps and CW regimes, should lead to realistic quantum relay schemes, namely by offering a reduced-constraint solution for the synchronization issue, compared to the fs regime, when two completely independent pump lasers are employed. Taking advantage of very high efficiency sources associated with high-end telecom components, such as FBG filters, was of utmost importance for all the realizations discussed above so as to ensure simultaneously a high degree of indistinguishability and a high overall brightness (see Ref.~\cite{Aboussouan_dipps_2010} for a detailed comparison).
Eventually, the use of the fs regime method comes with two drawbacks. On one hand, it requires a proper synchronization of the two remote sources to around 10\,fs accuracy, as mentioned above. If the electronic synchronization of two fs sources has been achieved in the laboratory~\cite{Kaltenbaek_Inter_Indep_2009}, this would be difficult in a setting where the sources are far away from each other. On the other hand, the major downside with such a regime of operation lies in the need for an accurate matching of the path lengths from both sources to the BSM apparatus, \textit{i.e.}, to within the coherence length of the interfering, or inner, photons (see Fig.~\ref{Fig_Qrelay})~\cite{Weihs_Synchro_2007}. Conversely, the main advantage of the ps and CW schemes amounts to having higher coherence time/length paired photons, thus reducing the need for path length stabilization.

\section{Up-conversion detectors and quantum transposition operations based on coherent wavelength conversion}
\label{Sec_Up_unitary}

In this section we discuss quantum optics and communication experiments based on other nonlinear optical processes than SPDC which is, as discussed in Section~\ref{Sec_sources}, widely used to produce entangled photon-pairs. In particular, we will consider sum and difference frequency generation (SFG - DFG) processes.
SFG, or up-conversion, has been employed for converting single photon wavelengths from the telecommunication bands (1310 and 1550\,nm) to the visible band (below 800\,nm), making it possible to use Silicon avalanche photodiodes (Si-APDs) that are not sensitive at telecom wavelengths~\cite{Albota_SFG_2004,Roussev_UpDet_2004,Vandevender_SFG_2004,Langrock_UpDet_2005,Diamanti_QKD_up_2005,Thew_GHzQKD_2006,Temporao_MID_06,Pan_SFG_2006,Thew_Tunable_SFG_2008,Dong_SFG_2008,tournier_up_2009}. In this case, advantages are multiple since Si-APDs show much better performance than Indium-Gallium-Arsenide (InGaAs) APDs usually employed for telecom detection~\cite{Stucki_Photon_counting_QKD_InGaAs_2001,Nakemata_SPAD_2002,Thew_InGaAs_2007}.

SFG, and more recently DFG, have also been investigated to perform quantum transposition operations when it is necessary to coherently transfer qubits carried by a photon at a given wavelength to another wavelength. Here by ``coherently" we understand that the quantum superpositions of states involved in the single or entangled qubits have to be maintained with a near unity fidelity during the nonlinear process. These operations, or \textit{quantum interfaces}, are of particular interest for connecting different nodes of a quantum network. Basically, such networks are made both of distribution telecom quantum channels along which photonic qubits travel~\cite{weihs_photonic_2001}, and for example, of atomic or ionic ensembles having transitions in the visible range and dedicated to storing and processing the qubits~\cite{Julsgard_QMem_2004,Langer_LongLivedQM_2005,chaneliere_storage_2005,lvovsky_optical_2009,tittel_PEQMSS_2009,reim_towards_2010}. Such quantum interfaces should allow transferring the qubits from one photonic carrier to another, \textit{i.e.}, provide a wavelength adaptation between photons and atomic transitions, while maintaining the quantum coherence of the initial state~\cite{Giorgi_FrequencyHopping_2003,Tanz_Interface_2005,curtz_CFDC_2010,takesue_SPFDC_2010}. Without being exhaustive, note that quantum teleportation can also provide coherent qubit wavelength transposition, as demonstrated in Ref.~\cite{Marcikic_Qtele_2003} between two photons lying in telecom bands (1310 to 1550\,nm), and proposed in Ref.~\cite{Chaneliere_Qtele_2006} between a telecom photon (1530\,nm)  and a rubidium cold atomic ensemble (780\,nm), respectively. Finally note that four-wave mixing can also be used for such coherent wavelength conversions. This has recently been demonstrated using a cold atomic ensemble~\cite{Radnaev_AtomInterface_2010}.

In any case, the role played by IO technologies, especially when based on PPLN waveguides, has been crucial for all the schemes considered below since they provided outstanding wavelength conversion probabilities, but also, thanks to better integration technologies, improved propagation loss figures.

\subsection{Up-conversion detectors based on integrated optics for telecom single photons}
\label{Sec_Up_det}

\subsubsection{A bit of history: detecting NIR and MIR photons}
The idea of up-conversion detectors, \textit{i.e.}, converting signal wavelengths, by means of nonlinear optics and dedicated pump lasers, to other wavelengths where better detectors are available is not new, and was proposed far before quantum communication, and related detection problems, appeared. In 1959, before the laser was demonstrated and the principles of nonlinear optics published, N. Bloembergen already discussed an ``infrared quantum counter'' based on optically pumped up-conversion in four-level ion-doped medium associated with a photomultiplier~\cite{Bloembergen_Qcounters_1959}. Without being exhaustive, the literature already reported, at the end of the 60's, quite complex up-conversion detectors based on nonlinear optics. Midwinter \textit{et al.} took advantage of a lithium niobate bulk crystal~\cite{Midwinter_SFG_1967} for detecting NIR photons while Anderson \textit{et al.} utilized state-of-the art IO technology with GaAs waveguides for detecting MIR photons~\cite{Anderson_SFG_1969}. Note that more recently, an MIR up-conversion detector has been developed by the Geneva group using PPLN to enable free-space quantum cryptography at 4.06\,$\mu$m wavelength~\cite{Temporao_MID_06}. 

\subsubsection{Why use up-conversion detectors for quantum communication ?} 
In the case of quantum communication over long distances, optical fibers are quite ideal channels for transporting photonic qubits since they show low propagation losses (0.17 and 0.3\,dB/km at 1550 and 1310\,nm, respectively). In addition, telecom wavelength based quantum communication demonstrations are advantageously enabled by the technological advances achieved over the past twenty years in classical telecommunication networks.
Unfortunately, the easiest way to detect telecom single photons, \textit{i.e.}, at 1310 and 1550\,nm, is to use InGaAs-APDs that, while well adapted for classical communication, show relatively poor performance in the single photon regime, particularly in terms of quantum detection efficiency and dark count probability. Table~\ref{Table_Si_InGaAs} summarizes the large difference of performance between InGaAs and Si-APDs, that operate in the visible range of wavelength, for single photon detection.

\begin{table*}
  \centering
  \caption{Compared performance of standard InGaAs~\cite{QKD_commercial} and Si-APDs~\cite{QKD_commercial,QKD_commercial_cova} for single photon detection as well as custom InGaAs-APDs~\cite{QKD_commercial,Namekata_HighSpeed_InGaAs_2006,Yuan_HighSpeed_InGaAs_2007,Thew_InGaAs_2007,Namekata_HighSpeed_InGaAs_2009,Dixon_InGaAs_shortDT_2009,Zhang_fast_10}. In addition to a significant difference in the quantum efficiencies (QE), one should note at least the one order of magnitude difference in the dark count (DC) probabilities, the large difference of the maximum achievable counting rates (CR), and the different operation regimes, \textit{i.e.}, free running or gated modes. Due to their high dark count probability, current commercial InGaAs-APDs are normally operated in a gated mode, meaning that the arrival times of the photons have to be known in advance. Under a continuous regime of operation, the photons' creation times remain unknown, implying that free-running detectors have to be employed. On the other hand, silicon-based detectors can be free running and, depending on the size of their active zone, can provide a timing jitter as short as 50\,ps~\cite{QKD_commercial,QKD_commercial_cova}, however with a maximum efficiency reduced to $\sim$30\%. Although Si-APDs show high performance, they cannot be employed for detecting telecom single photons unless a wavelength adaptation of the incoming single photons is performed, for instance by means of nonlinear up-conversion. Custom InGaAs detectors, that are now capable of working either in free running mode~\cite{QKD_commercial,Thew_InGaAs_2007} or in fast gated mode~\cite{Namekata_HighSpeed_InGaAs_2006,Yuan_HighSpeed_InGaAs_2007,Namekata_HighSpeed_InGaAs_2009,Dixon_InGaAs_shortDT_2009,Zhang_fast_10}, are the first glimmerings of performant telecom wavelength single photon APDs. Note that free running InGaAs detectors reached the market very recently~\cite{QKD_commercial}. Com.: Commercial; Cust.: Custom.}
  \label{Table_Si_InGaAs}
  \begin{tabular}{@{}lcccc@{}}
    \hline
   APD type & Com. Si~\cite{QKD_commercial,QKD_commercial_cova} & Com. InGaAs~\cite{QKD_commercial} & Cust. InGaAs~\cite{Yuan_HighSpeed_InGaAs_2007} & Cust. InGaAs~\cite{QKD_commercial,Thew_InGaAs_2007} \\
    \hline
    Max. QE & $\sim 0.6$ &  0.1-0.2 & 0.11 &  0.1 \\
    DC prob. (ns$^{-1}$) &  10$^{-8}$&  10$^{-5}$-10$^{-6}$ &  2.3$\cdot$10$^{-6}$ & $\sim$10$^{-5}$ \\
    Max. CR (Mcounts/s) & 8 &  0.15 & 100 & N/A \\
    Oper. mode & free running &  gated ($\sim$2\,MHz) & fast gating (1.25\,GHz) & free running \\
    Det. range (nm) & 400-800 & 1100-1600 & 1100-1600 & 900-1600 \\
    Timing jitter (ps) & 50-500 & 500 & 55 & 100 \\
    \hline
  \end{tabular}
\end{table*}

As a consequence, with a basic quantum channel operating at 1550\,nm involving a single photon source (see Section~\ref{Sec_sources}) linked to InGaAs-APDs with a fiber, one can easily understand, taking into account the losses in the fiber, that the noise figure of the detector represents the main limiting factor in terms of overall SNR and therefore of the maximum achievable distance. In other words, using InGaAs-APDs enables quantum link distances on the order of 150\,km while Si-APDs, provided wavelength conversion is applied, would extend the distance to about 250\,km~\cite{gisin_QKD_2002,QKD_commercial,Collins_QRelays_2005,Nakemata_SPAD_2002,Thew_InGaAs_2007}.
Increasing quantum communication distances therefore amounts to improving the detection process, especially reducing the noise figure, since optical fiber technology has reached its theoretical limits in terms of propagation losses. Quantum communication scientists had the idea of hybrid up-conversion detection which seems to be a good alternative since it allows, at least theoretically, to translate photons from the telecom range to photons in the visible range via the association of the SFG process and permit the use of high performance Si-APDs~\cite{Albota_SFG_2004,Roussev_UpDet_2004,Vandevender_SFG_2004,Langrock_UpDet_2005,Diamanti_QKD_up_2005,Pan_SFG_2006,Temporao_MID_06,Dong_SFG_2008,Thew_GHzQKD_2006,Thew_Tunable_SFG_2008,tournier_up_2009}.

Note that another type of detector based on superconducting nano-wires, has been developed over the last few years. Such detectors have shown an efficiency of up to 50\% at the telecom wavelengths for a noise comparable to that of the Si-APDs~\cite{Verevkin_SupraDet_04,Engel_SupraDet_04,Hu_sspd_11}. However, the implementation of these devices is still challenging, both in terms of manufacturing and reproducibility, and complicated, both in terms of handling and operational expenses (liquid helium cooling).

\subsubsection{State of the art using integrated optics technologies}
\begin{figure}
\centering%
\includegraphics[width=\columnwidth]{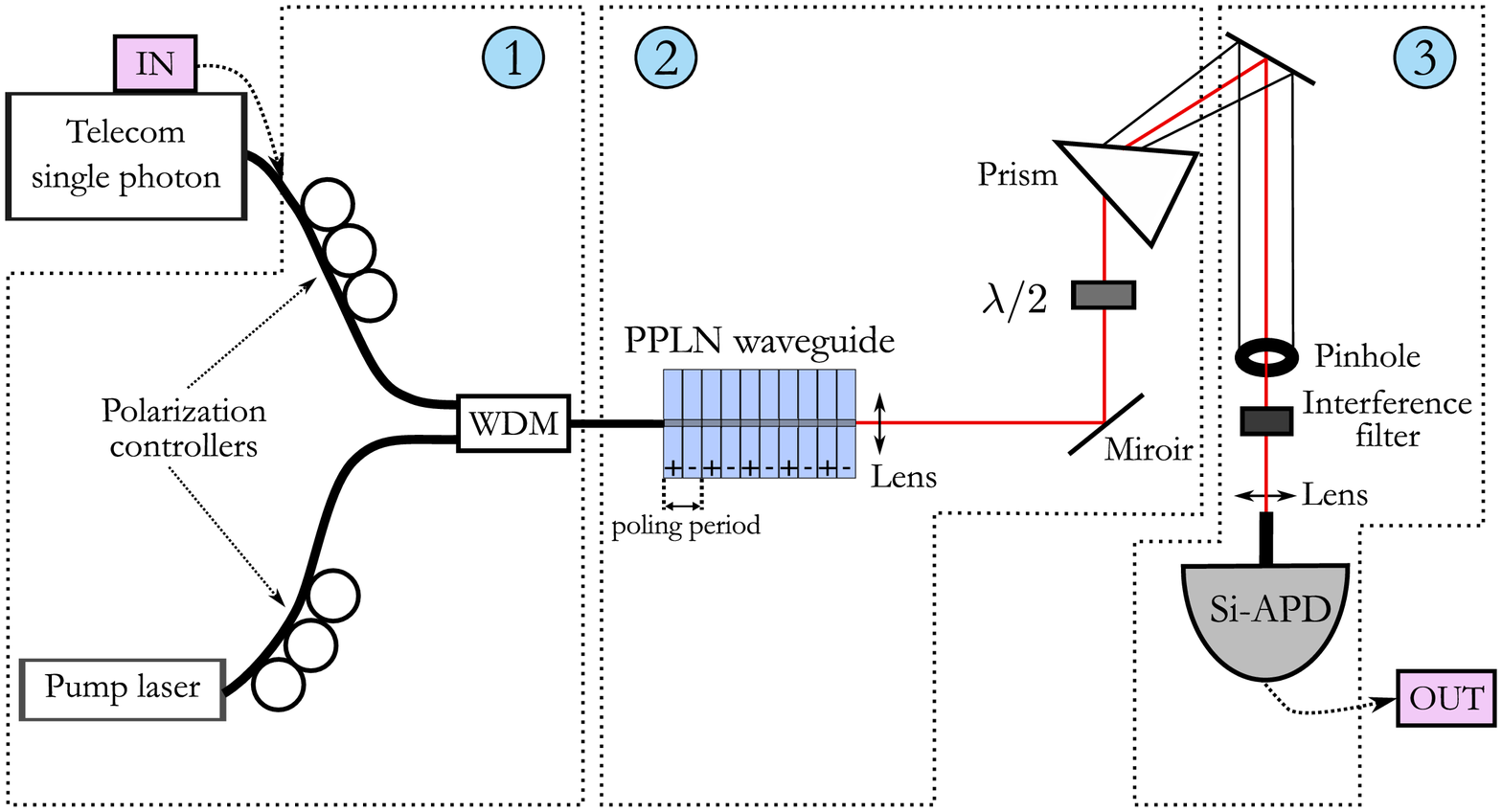}
\caption{\label{Fig_Hybrid}General scheme for hybrid detection based on a single photon source at a telecom wavelength, a pump laser, and an IO wavelength converter. (1) All the components are ``ideally" fibered. The incoming single photons can lie at 1550\,nm, and the pump laser can be a standard diode laser emitting at 980\,nm. The signal and pump wavelengths are then mixed through a WDM and enter, for instance, a PPLN waveguide. Depending on the waveguide fabrication technology (PE or titanium in-diffusion), polarization directions of the two input wavelengths might be adjusted thanks to polarization controllers. (2) The nonlinear crystal is temperature controlled so as to reach the necessary QPM condition. (3) The light beam at the waveguide output contains SFG photons (e.g. at 600\,nm with 1550 and 980\,nm as inputs), residual pump photons and non-converted signal photons. This beam is collimated and then dispersed by a prism in order to separate the different wavelengths. Then, the SFG photons pass through a narrow filter centered at the SFG wavelength, allowing for maximum elimination of the remaining pump photons. The photons are then collected by an optical fiber and sent to an Si-APD. The number of possibilities is large regarding pump and signal wavelengths, as well as regarding the employed nonlinear crystal.
Although PPLN waveguides have been widely employed, note that there are also many realizations based on bulk crystals, such as PPLN or PPKTP. Good examples are given in Refs.~\cite{Albota_SFG_2004,Vandevender_SFG_2004,Pan_SFG_2006,Temporao_MID_06,Dong_SFG_2008}. Figure extracted from Ref.~\cite{tournier_up_2009}.}
\end{figure}

Up-conversion, or hybrid, detectors have therefore been investigated in quantum communication experiments to replace InGaAs-APDs by Si-APDs (see the general scheme in Fig.~\ref{Fig_Hybrid}). Those detectors are based on the supposedly noise-free process of sum-frequency generation, or SFG, using a second order nonlinear crystal. Powered by an intense pump laser, this process permits converting, with a certain probability, the single photons at telecom wavelengths to the visible range where Si-APDs operate (see Table~\ref{Table_Si_InGaAs}). To date, the literature reports up-conversion detectors having main features, \textit{i.e.}, efficiency and noise figures, comparable to that of the best commercially available InGaAs-APDs. However, in all the realizations, a pump-induced noise was always observed when the pump wavelength is shorter than the signal to be converted. Interestingly, the noise figure was initially expected to be as low as the dark count level of the Si-APDs. Although this additional noise affects their overall performance, up-conversion detectors can replace InGaAs-APDs in various long-distance and high-speed quantum cryptography schemes since they offer a continuous operation regime mode instead of the gated mode necessary for standard InGaAs-APDs. This permits the possibility of much higher counting rates, and possibly much shorter timing jitter depending on the employed Si-APD (see Table~\ref{Table_Si_InGaAs})~\cite{QKD_commercial,Diamanti_QKD_up_2005,Thew_GHzQKD_2006}. 

Many hybrid detectors for quantum communication experiments operating at the telecom wavelengths of 1310 and 1550\,nm have been reported in the literature. A detailed review on the recent advances in the field is given in Ref.~\cite{tournier_up_2009}. These detectors are based on various pump schemes and different nonlinear crystals. Without being exhaustive, we shall present some of the more important realizations by grouping them in terms of combinations of crystal/pump laser.
To understand the figures presented in the following, we may recall that the overall efficiency of a hybrid detector includes the up-conversion probability of the SFG process which depends on the pump power, the detection efficiency of the employed Si-APD, as well as the transmission of the filtering stage.

Even if not based on IO technologies, a detector developed at MIT in 2004 was perhaps the first in this area~\cite{Albota_SFG_2004}. It was based on a laser at 1064\,nm and pumping a PPLN bulk crystal placed in a cavity so as to enhance the up-conversion efficiency of photons at 1550\,nm to photons at 630\,nm. The same configuration was used at the University of Urbana Champaign, again in 2004~\cite{Vandevender_SFG_2004}, and at the University of Shanghai in 2006~\cite{Pan_SFG_2006}. These three detectors exhibited very high internal up-conversion efficiency (about 80\%) however with a high noise level, of about 4$\cdot$10$^5$\,counts/s.\\
Using IO nonlinear crystal solutions, \textit{i.e.}, mainly based on PPLN waveguides, excellent performances have been achieved:
\begin{itemize}
\item In 2004, a realization from the University of Stanford based on a reverse PE:PPLN waveguide reported a global detection efficiency of 41\% without specifying the noise level~\cite{Roussev_UpDet_2004}.
\item In 2005 the same group, in collaboration with the NTT Japanese research group, realized a more complete characterization, notably by providing conversion efficiency curves and the corresponding noise~\cite{Langrock_UpDet_2005} to carry out quantum cryptography experiments at 1550\,nm~\cite{Diamanti_QKD_up_2005}. For this experiment, the authors used pumps at 1310 and 1550\,nm to convert single photons at 1550 and at 1310\,nm, respectively, both to the final wavelength 710\,nm. When the pump was at 1310\,nm, the detector exhibited a maximum global detection efficiency of 45\% and a noise level of 8$\cdot$10$^5$\,counts/s, while when the pump wavelength was 1550\,nm, the efficiency was about 40\% for a noise level of 2$\cdot$10$^4$\,counts/s. In other words, they obtained almost the same efficiency with a 40 times reduction of the noise counts in the latter case. Here, one can appreciate the importance of the pump wavelength's being larger or smaller than the signal wavelength, on the noise level induced by the pump in the nonlinear process.
\item In 2006, the Geneva Group performed a complete characterization of a hybrid detector combining a pump at 980\,nm and a PE:PPLN waveguide, for the conversion of photons at 1550\,nm to photons at 600\,nm. In this configuration, the authors obtained an overall efficiency of about 5\% for a noise level of 5$\cdot$10$^4$\,counts/s~\cite{Thew_GHzQKD_2006}. In the Geneva experiment, the pump power was reduced so as to minimize the noise so that a high-speed quantum cryptography experiment could be conducted which would otherwise not have been possible without the use of hybrid detectors. This is why the global efficiency and noise are an order of magnitude smaller when compared to those achieved by the MIT group~\cite{Langrock_UpDet_2005,Diamanti_QKD_up_2005}. However, the ratio overall efficiency over noise level is the same for both realizations~\cite{Langrock_UpDet_2005,Thew_GHzQKD_2006}.
\item More recently, in 2008, the Geneva group developed a hybrid detector based on the same configuration as before but with a pump laser which can be tuned by utilizing a Bragg grating~\cite{Thew_Tunable_SFG_2008}. The hybrid detector was tunable over a wavelength range of 5\,nm and showed features that are almost the same as those mentioned before~\cite{Thew_GHzQKD_2006}.
\end{itemize}

In conclusion, additional noise appears in hybrid detectors when the pump wavelength is shorter than the signal to be converted. As we will see in the next section, the Geneva group, in collaboration with our group, implemented a coherent qubit wavelength-transfer experiment which was essentially noise-free since entanglement has been shown to be preserved during the up-conversion process~\cite{Tanz_Interface_2005}. As in Ref.~\cite{Langrock_UpDet_2005}, the pump wavelength was around 1550\,nm while the signal photons were at 1310\,nm. Ref.~\cite{tournier_up_2009}, in which a systematic study of this noise has been carried out using a PPLN waveguide and a pump at 980\,nm wavelength, showed that the process of SPDC was responsible for this background noise. This process involves one pump photon at 980\,nm that splits into energy-correlated photons, with one at 1550\,nm which corresponds to the signal, and one at 2660\,nm. The noise is then due to the up-conversion of these ``noise photons" at 1550\,nm thanks to the presence of the pump having a shorter wavelength. Unfortunately, the resulting ``noise converted photons" cannot be filtered from the regular photons since they have exactly the same wavelength of 600\,nm.

Today, hybrid detectors for telecom photons show efficiencies comparable to those of InGaAs-APDs but can be troubled by noise levels as explained above. Nevertheless, the real progress lies in the possibility of detecting single photons in a continuous regime, as Si-APD's do, instead of a gated mode necessary for limiting the dark count level in InGaAs-APD's. In addition, the detection rate can be much higher than that permitted by standard commercially available InGaAs-APDs (see Table~\ref{Table_Si_InGaAs}). It must be noted however that InGaAs-APD's technology has been highly improved during the past years since the realization of custom devices, working in either a free running regime or in a fast gating mode, have been demonstrated~\cite{Namekata_HighSpeed_InGaAs_2006,Yuan_HighSpeed_InGaAs_2007,Thew_InGaAs_2007,Namekata_HighSpeed_InGaAs_2009,Dixon_InGaAs_shortDT_2009,Zhang_fast_10,QKD_commercial} (see Table~\ref{Table_Si_InGaAs}). These define new references in the field, even if not yet commercially available.

\subsection{Quantum transposition operations based on coherent wavelength conversion}

As mentioned earlier, quantum entanglement is widely recognized to be at the heart of quantum physics, with all its counter intuitive features. The new science of quantum information treats entanglement as a resource for quantum teleportation and relays, which is essential for the coming age of quantum technology. Yet, to be of any practical value, it has to be possible to manipulate it. Accordingly, a growing community of experimental physicists are working with entanglement transposition, especially in view of the connection of different users on a quantum network consisting of quantum channels along which photonic qubits travel~\cite{weihs_photonic_2001} and atomic ensembles are used to store and process the qubits~\cite{Julsgard_QMem_2004,Langer_LongLivedQM_2005}. It is thus timely to develop quantum information interfaces that allow transferring the qubits from one type of carrier to another, \textit{i.e.}, in the case of photonic carriers, able to provide a wavelength adaptation to atomic levels while preserving quantum coherence of the initial state~\cite{Giorgi_FrequencyHopping_2003,Tanz_Interface_2005,curtz_CFDC_2010,takesue_SPFDC_2010}.

In 2005, the coherent transfer of a qubit from a photon at the telecom wavelength of 1312\,nm to another photon at 712\,nm, \textit{i.e.}, close to that of alkaline atomic transitions~\cite{Julsgard_QMem_2004,Langer_LongLivedQM_2005,Chaneliere_RemoteAtomEnt_06,Pan_QMRb_2008,Pan_QMRb_2009}, was demonstrated. This operation has been performed in a general way. The initial information-carrying photon (at 1312\,nm), entangled in energy and time with another photon at 1555\,nm, was up-converted to 712\,nm. It was then verified that the transfer did preserve the quantum coherence by testing the entanglement between the received photon at 712\,nm and the remaining photon at 1555\,nm. The entanglement was found to be almost perfect, despite the fact that these two photons did not satisfy the energy conservation rule of Eq.~\ref{Eq_conservation} anymore~\cite{Tanz_Interface_2005}. In other words, a coherent transfer from one photon to another at a different wavelength does not alter the original quality of entanglement.
\begin{figure}[h!]
\centering%
\includegraphics[width=\columnwidth]{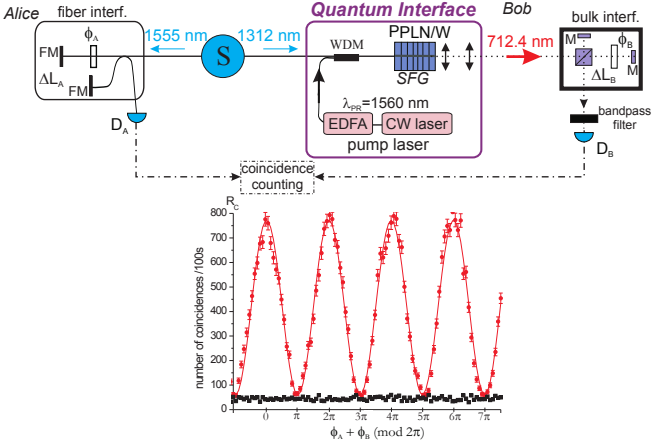}
\caption{\label{Fig_Interface}A photonic quantum interface dedicated to transferring coherently single qubits from photons at 1310\,nm, initially entangled with photons at 1550\,nm, to photons at 710\,nm, and to demonstrating that entanglement is preserved through the up-conversion process. First, the source (S) produces, by parametric down conversion in a PPLN waveguide (not represented), pairs of energy-time entangled photons whose wavelengths are centered at 1555 and 1312\,nm, respectively. By means of telecommunications optical fibers, these photons are sent to Alice and Bob, respectively. Next, the qubit transfer is performed at Bob's location from a photon at 1312\,nm to a photon at 712\,nm using the SFG process in a PE:PPLN waveguide (PPLN/W). This crystal is pumped by a CW, 700\,mW, and high coherence laser working at 1560\,nm. The success of the transfer is tested by measuring the quality of the final entanglement between the newly created 712\,nm photon and Alice's 1555\,nm photon using two unbalanced Michelson interferometers and a coincidence counting technique between detectors D$_A$ and D$_B$. The graph shows the interference pattern obtained for the coincidence rate as a function of the sum of the phases acquired in the interferometers ($\phi_A + \phi_B$). The corresponding visibility is greater than 97\%. Compared to the 98\% obtained with the initial entanglement (curve not represented), the quantum fidelity of this transfer is greater than 99\%. Figure extracted from Ref.~\cite{Tanz_Interface_2005}.}
\end{figure}

This has been achieved using up-conversion, \textit{i.e.}, by mixing the initial 1312\,nm single photon with a highly coherent pump laser at 1560\,nm in a PE:PPLN waveguide, as depicted in Fig.~\ref{Fig_Interface}. The probability of a successful up-conversion was $\sim$5\%, including the losses due to phase-matching in the nonlinear waveguide and the addition of an interference filter centred on the resulting wavelength. The received single photon at 712\,nm, filtered out of the huge flow of pump photons, and the remaining 1555\,nm photon was then analyzed using two unbalanced Michelson interferometers in the Franson conformation~\cite{Franson_Bell_1989} (\textit{i.e.}, the usual setup for testing energy-time or time-bin entanglement). From the high quality of the resulting two-photon interference, we conclude that the fidelity of this quantum information transfer was excellent, higher than 99\%. Here, the fidelity has been evaluated by using the comparison between the visibilities of the interference patterns obtained with and without the quantum transfer (more details are given in the caption of Fig.~\ref{Fig_Interface}). We emphasize that this experimental result is very encouraging for quantum information science and that this improvement of the mastering of entanglement has been enabled by IO technology.

More recently, two new quantum interfaces have been demonstrated, utilizing the process of single photon down-conversion, or DFG, one by the Geneva group~\cite{curtz_CFDC_2010}, and the other by a NTT Japanese research group~\cite{takesue_SPFDC_2010}.
The idea here is to proceed to the exact inverse operation to that discussed just above, \textit{i.e.}, transferring coherently qubits initially carried by photons in the visible range back to the telecom range. In other words, both papers, dealing with quite the same experimental procedures, propose a solution towards re-encoding qubits onto telecom photons after the read-out of a quantum memory based on alkaline atomic ensembles~\cite{Julsgard_QMem_2004,Chaneliere_RemoteAtomEnt_06,Pan_QMRb_2008,Pan_QMRb_2009}. 
\begin{figure}[h!]
\centering%
\includegraphics[width=\columnwidth]{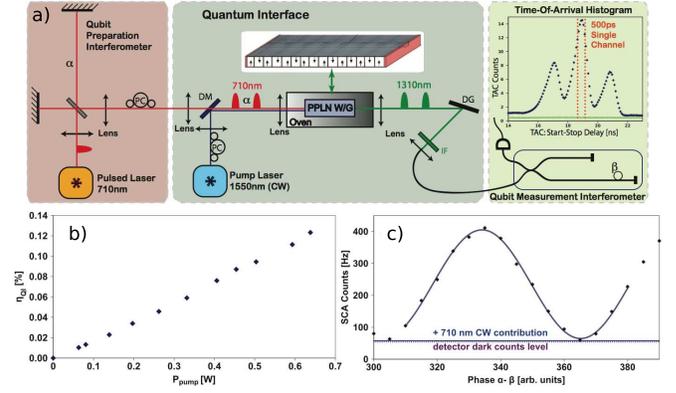}
\caption{\label{Fig_Interface_New}A photonic quantum interface dedicated to transferring coherently single qubits from photons at 710\,nm to photons at 1310\,nm by means of DFG. (a) Experimental set-up to characterize the coherence of the quantum interface involving an attenuated pulsed laser mimicking a single photon source at 1310\,nm, a preparation bulk interferometer encoding time-bin qubits on the 710\,nm photons, a pump laser at 1550\,nm, a PE:PPLN waveguide enabling single photon down-conversion (or DFG), a fiber interferometer for time-bin qubit analysis, and a dedicated electronics that enables recording the difference in arrival time between the pulsed laser source and the detection (inset, topright). (b)  Quantum interface efficiency as a function of the 1550\,nm pump power in the waveguide. (c) Count rates for interfering events detected by a suitable AND-gate for an average number of prepared single photons reaching unity. As the phase of the fiber interferometer is scanned, a net (raw) interference visibility of 96\% (84\%) is obtained. Figures extracted from Ref.~\cite{curtz_CFDC_2010}.}
\end{figure}
Both experiments amount to generating single photons from an attenuated pulsed laser and to encoding them as time-bin qubits thanks an interferometer. After the single photon down-conversion is processed in a PE:PPLN waveguide, qubit analysis is made using either a fiber interferometer (see Fig.~\ref{Fig_Interface_New}(a) and Ref.~\cite{curtz_CFDC_2010}) or a PLC interferometer (see Ref.~\cite{takesue_SPFDC_2010}). The obtained results, in terms of interfering time-bin events of the form $\frac{1}{\sqrt{2}}\left(\ket{short}_p\ket{long}_a + \ket{long}_p\ket{short}_a\right)$, where $short$ and $long$ denote paths in the preparation ($p$) and analysis ($a$) interferometers, respectively, are of the same order, \textit{i.e.}, showing a net visibility of almost 96\% for the Geneva group and of 94\% for the NTT group.

Note that the quantum interfaces discussed above are not limited to the specific wavelengths chosen. Indeed, suitable modifications of phase-matching conditions and pump wavelengths enable tuning the result of the up-conversion process to any desired wavelength. More precisely, periodically poled waveguiding structures, as described above (see Sec.~\ref{Sec_NLO_waveguide}), provide the experimentalists with a broad range of accessible wavelengths by changing the poling period and the pump wavelength accordingly~\cite{tanzilli_ppln_2002}. For instance, having an original qubit carried by a photon at 1550\,nm and a pump laser at 1570\,nm would lead to a transferred qubit carried by a photon at the wavelength of 780\,nm, \textit{i.e.}, that of rubidium atomic transitions~\cite{Chaneliere_RemoteAtomEnt_06,Pan_QMRb_2008,Pan_QMRb_2009}.
On the other hand, such IO components yield very high up- and down-conversion efficiencies, as widely discussed throughout this paper. They permit using modest reservoir powers to achieve reasonable qubit transfer probabilities. In the case of applications requiring very narrow photon bandwidths, for instance when transferring photonic qubits to atoms~\cite{Shapiro_LongDistTele_2001,Julsgard_QMem_2004,Langer_LongLivedQM_2005,Chaneliere_RemoteAtomEnt_06,Pan_QMRb_2008,Pan_QMRb_2009}, bright down-converters make very narrow spectral filtering possible while maintaining a high photon-pair flux, with reasonable pump powers as described in Sec.~\ref{Sec_sources} and in Refs.~\cite{Halder_Ent_Indep_2007,Aboussouan_dipps_2010}.

However, although the current technology of Bragg grating filters allows having photon bandwidths as narrow as a few pm, this is still a limiting factor when the purpose is to map these photons to atomic transitions. Depending on the nature of the considered atomic ensembles, the bandwidths are too large by one to three orders of magnitude.
Even if the past ten years have seen the development of single and entangled photon sources of higher and higher brilliance~\footnote{This is the number of photon-pairs created per nm and per mW of pump power.}, it is very important to note that efforts are also coming from the atomic side since physicists are now trying to develop broadband mapping of single photons as described in recent proposals~\cite{Walmsley_BroadMapping_2007,reim_towards_2010,Saglamyurek_BroadbandWQM_2011}.		

To conclude this section, note that photonic and atomic qubit mapping is of great interest since it would facilitate the building of quantum networks and computers.

\section{Integrated optical waveguides for quantum memories}
\label{Sec_QM}

Improving the performance of long distance quantum communication, in view of building real quantum networks, necessitates the development of quantum memories. Such devices can be employed  for realizing deterministic single photon-sources~\cite{Matsukevich_DSPS_2006,Chen_DSPS_2006}, for turning quantum relays to much more efficient quantum repeaters~\cite{Simon_QR_2007,Simon_multiQR_2010}, and for further enabling quantum computation~\cite{Shor_DecoQC_1996,Knill_QECC_1997,Kok_QC_2007,knill_QC_2010}. A quantum repeater associates a memory function with a quantum relay. Thanks to this the desired qubit can be recalled when it is needed for further operations.

Quantum memories figures of merit are typically the storage time, the storage efficiency, the fidelity and the spectral acceptance bandwidth~\cite{tittel_PEQMSS_2009}. The efficiency represents the probability of absorption/re-emission in the right temporal and spatial modes. The fidelity corresponds to the overlap between the output qubit state and the input one. Note that a broader  spectral acceptance bandwidth permits increasing the speed of quantum communication links.

In the race towards quantum memory realizations, many quantum storage protocols have been investigated and associated with a variety of possible devices. We find in the literature, among others, solutions based on alkaline cold atom ensembles associated with Electromagnetically Induced Transparency (EIT)~\cite{chaneliere_storage_2005} or with the DLCZ protocol~\cite{duan_long-distance_2001,Sangouard_DLCZRMP_2011}, atomic vapors associated with off-resonant two-photon transitions~\cite{reim_towards_2010}, ion-doped crystals associated with photon echo or atomic frequency combs~\cite{de_riedmatten_ss_2008,chaneliere_light_2010}, \textit{etc.} The reader can find very pertinent review papers in this field in Refs.~\cite{lvovsky_optical_2009,tittel_PEQMSS_2009,Hammerer_interface_10}. 

These combinations offer various interaction wavelengths and linewidths, ranging from the visible to the telecom band, and from a few MHz to almost 5 GHz, respectively. In any case, both the wavelength and bandwidth of paired photons from sources as described in Section~\ref{Sec_sources} must be adapted to the considered storage device and protocol. In particular, an additional filtering stage has to be employed in order to reduce the bandwidth by at least two orders of magnitude.
This can be done by using a Fabry-Perot cavity, provided the pump laser power is increased to some extend so as to recover an acceptable pair production rate within the limit of the considered experiment regarding the created mean number of pairs (should be $<$1) per relevant time window~\cite{halder_high_2008,pomarico_waveguide_2009}. This is where the advantage of integrated nonlinear optics schemes lies in terms of efficiency. While a few hundreds of mW of pump power should be enough, one has to pay extra attention to double pair emission within the time detection window that could lead to a decrease of the entanglement quality.

Quantum memory solutions based on ion-doped crystals, working at cryogenic temperatures (several K), have been investigated using IO configurations. Historically, the first implementation was proposed by Hastings-Simon and collaborators, where they studied the Stark effect in Er$^{3+}$-doped Ti:LiNbO$_3$ waveguides~\cite{staudt_interference_2007,staudt_fidelity_2007}, and proved the suitability of the scheme for solid state quantum memory protocols based on Controlled Reversible Inhomogeneous Broadening (CRIB)~\cite{Kraus_CRIB_2006,Gisin_PEQM_2007,Lauritzen_SSM_2010}, which is an extension of the photon echo technique showing better efficiency and fidelity. Storage times of 1 to $2\,\mu$s were demonstrated and a possibility of extending this to $6\,\mu$s discussed while the acceptance bandwidth is maintained at 125\,MHz. This solution is fully compatible with standard telecommunication networks since an operation wavelength of 1550\,nm can be directly mapped to the transition ($^{4}I_{15/2} \rightarrow ^{4}I_{13/2}$). Moreover, the Er$^{3+}$ ions are embedded in a IO waveguide enabling input and output fiber pigtails.

\begin{figure}
\centering%
\includegraphics[width=\columnwidth]{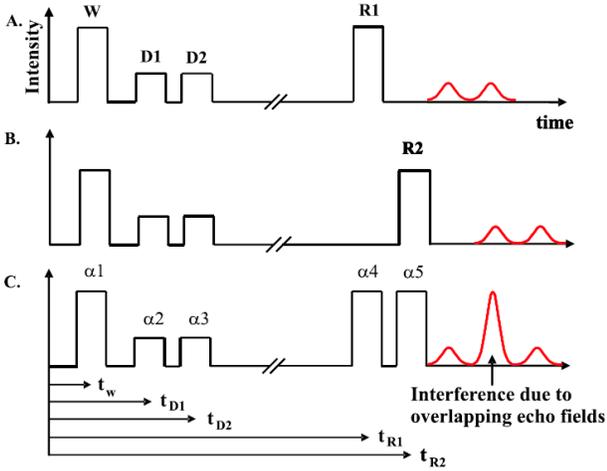}
\caption{\label{Fig_Memoire_pulse_sequence}Illustration of the pulse sequence used for time-bin pulses interference in a Ti:Er:LiNbO$_3$ waveguide memory. Figure extracted from Ref.~\cite{staudt_fidelity_2007}.}
\end{figure}

Two recent papers reported on the characterization of Ti:Er:LiNbO$_3$ waveguide devices operated with stimulated photon-echo, suitable for implementation of solid-state quantum memories. Details of the photon-echo operation principle can be found in~\cite{mitsunaga_time-domain_1992,tittel_PEQMSS_2009}. The first experiment demonstrated the possibility of storing the amplitude and phase of two subsequent coherent pulses is depicted in Fig.~\ref{Fig_Memoire_pulse_sequence}~\cite{staudt_fidelity_2007}.

Two pulses are sent to the memory after a strong write pulse. Then a read pulse is used to retrieve the stored information, the relative times between the read pulse and the echo emissions being the same as the relative times between the write and data pulses. To observe interference, two read pulses are sent to the memory with a time difference equivalent to the delay between the two data pulses. This pulse sequence configuration gives three echoes, but the second one is initiated by two contributions that are overlapping, \textit{i.e.}, the second data pulse emitted by the first read pulse and the first data pulse triggered by the second read pulse. An interference pattern is observed thanks to the modulation of the phase carried by either the data or the read pulse. In this case, the obtained visibilty reaches 100\%. This experiment demonstrates a proof of the phase and amplitude preservation of two subsequent coherent pulses and the possibility of adjusting the phase.
The second experiment, based on the same technology as before, demonstrated the coherent storage of the relative phase between two pulses in two independent quantum memories, each placed in an arm of a fibered Mach-Zehnder interferometer~\cite{staudt_interference_2007}. The phase in the interferometer could be modulated using a piezo-electric device, leading to an interference pattern showing a visibility of 90.5\%, and therefore proving phase preservation over the storage, and indistingishability of photon echoes. 

\begin{figure}
\centering%
\includegraphics[width=\columnwidth]{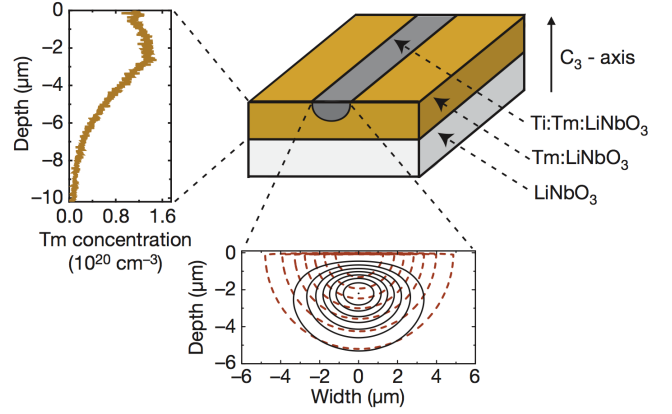}
\caption{\label{Fig_QM_waveguide}Illustration of the waveguide geometry for implementing a solid-state quantum memory, in which the different composition layers are depicted. The measured Tm concentration profile is given on the left and the calculated intensity distribution of the fundamental TM-mode at the 795\,nm wavelength is shown below. Iso-intensity lines are plotted corresponding to 100\%, 87.5\%, 75\% and so on of the maximum intensity. Figure extracted from Ref.~\cite{Saglamyurek_BroadbandWQM_2011}.}
\end{figure}

Very recently, a thulium-doped LiNbO$_3$ waveguide has been investigated as an alternative to the Ti:Er:LiNbO$_3$ through a Calgary/Paderborn collaboration~\cite{sinclair_spectroscopic_2010,Saglamyurek_BroadbandWQM_2011}. With such a device, shown in Fig.~\ref{Fig_QM_waveguide} the thulium (Tm$^{3+}$) transition of interest is at 795\,nm and the authors showed that the related transition has an optical coherence of 1.6\,$\mu s$ at a cooling temperature of $3\,K$. A storage time of up to 7\,ns was demonstrated. The authors showed the coherent storage of time-bin entangled photons with a photon-echo quantum memory protocol. Thanks to this Tm-doped waveguide, they showed a broadband storage photon bandwidth of 5\,GHz, which corresponds to an increase by a factor 50 when compared to earlier published results for solid-state quantum memory devices. This now compares well with systems based on atomic vapours, such as the ones demonstrated by the Oxford group~\cite{Walmsley_BroadMapping_2007,reim_towards_2010}.
Last but not least, they obtained an input/output entanglement fidelity or more than 95\%.
This realization emphasizes why solid-state quantum memory devices based on IO technology, and especially ion-doped Ti-indiffused waveguides, could play a leading role in future quantum memory development implementing photon-echo based protocols.

Although not based on a waveguide device, but published at the same time as the previously discussed realization, note that the Geneva group reported the demonstration of entanglement between a photon at a telecommunication wavelength (1338\,nm) and a single collective atomic excitation stored in a bulk crystal~\cite{Clausen_QSPECrystal_2011}. The quantum memory was made of a 1-cm-long Nd:Y$_2$SiO$_5$ crystal, \textit{i.e.}, impurity-doped with neodymium ions having a resonance at 883\,nm with good coherence properties and based on a photon-echo-type interaction using an atomic frequency comb (AFC). One photon at 883\,nm, from an energy-time entangled pair, has been mapped onto the crystal and then released into a well-defined spatial mode after a predetermined storage time while the complementary telecom photon was sent directly through a 50\,m fiber link to an analyser. Successful storage of entanglement in the crystal was proven by a violation of the Bell-Clauser-Horne-Shimony-Holt inequality~\cite{BCHSH_1969} by almost three standard deviations ($S = 2.64\pm0.23$). The storage time ranged from 25 to 200\,ns.

As was the case for the previous discussed demonstration, although operating at cryogenic temperatures, these results~\cite{Saglamyurek_BroadbandWQM_2011,Clausen_QSPECrystal_2011} represent an important step towards quantum communication technologies based on solid-state devices, in particular for building multiplexed quantum repeaters for long-distance quantum networks~\cite{Simon_QR_2007}.

\section{Integrated optical technologies for quantum computing based on linear optics}
\label{Sec_QComp}

Quantum Information Science (QIS) aims at harnessing quantum mechanical effects to develop quantum systems that will deliver significant improvements in information processing~\cite{Deutsch85,DiVincenzo95,Nielsen00}, such as exponentially faster factoring and simulations and quadratically faster data base searches.

However, a significant hurdle for systems that are more complex than ``simple quantum communication" is realizing interactions between two photons, in order to generate entangled states and to perform logic operations. Such interactions would require giant optical nonlinearities, stronger than those available in conventional nonlinear media. This leads to the consideration of EIT~\cite{Schmidt:96}, and atom-cavity systems~\cite{Turchette95}, although both of which present major technical challenges. In 2001, a major breakthrough known as the KLM (Knill-Laflamme-Milburn) scheme showed that scalable quantum computing is possible using only single-photon sources and detectors, and linear optical circuits~\cite{Knill01}. Since then, we have seen many experimental proofs-of-principle of two-qubit gates~\cite{O'Brien03,Pittman03,Takeuchi10}, as well as demonstrations of simple error-correcting codes~\cite{O'Brien05,Lu08}, and simple quantum algorithms~\cite{Lu07,Lanyon07}.

\begin{figure}
\centering%
\includegraphics[width=\columnwidth]{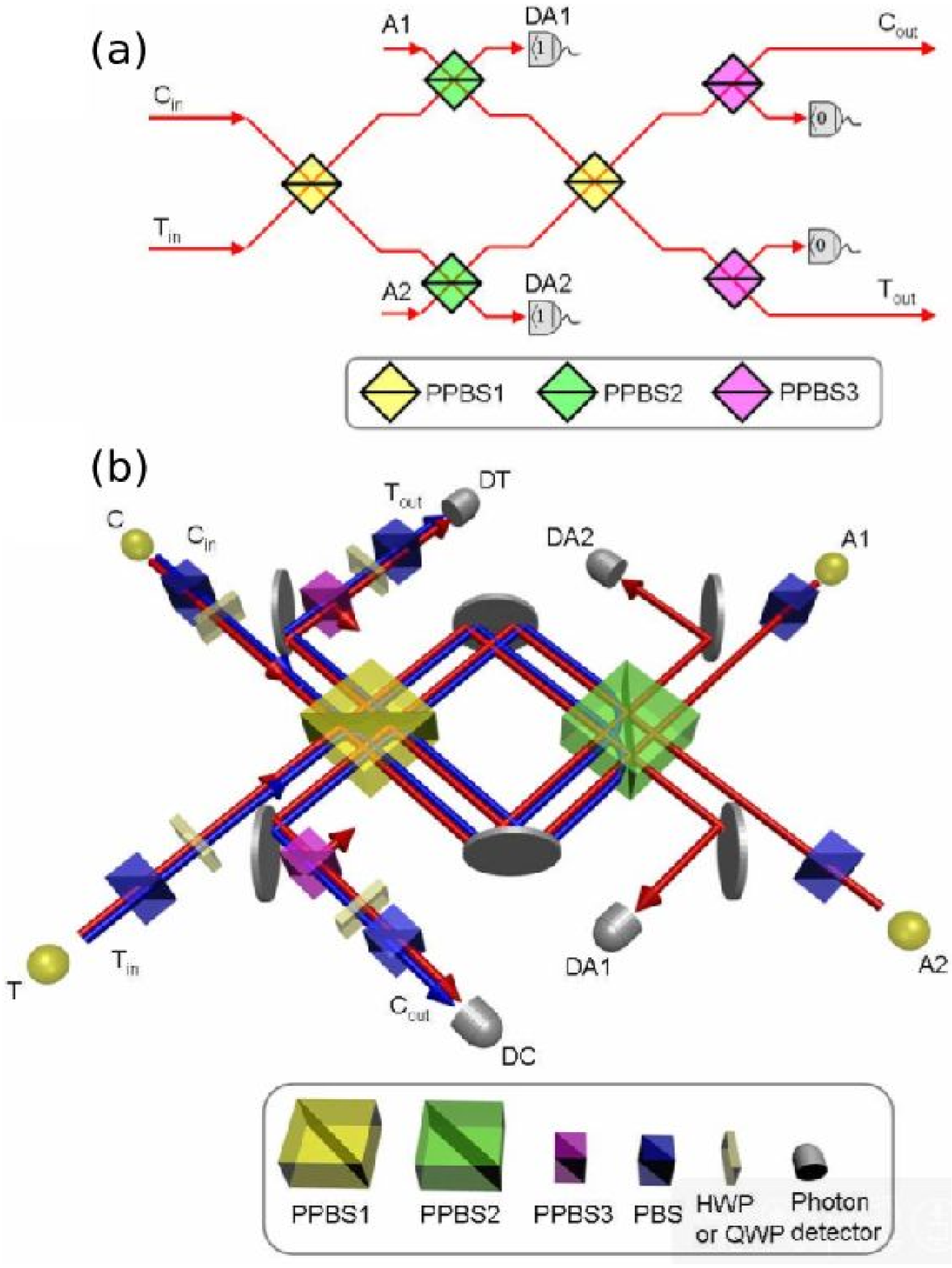}
\caption{Schematic of the KLM CNOT logic gate (a) and its folded realization with bulk optics (b). Figures extracted from Ref.~\cite{Takeuchi10}.\label{Fig_bulkTakeuchi}}
\end{figure}

Most of the demonstrations of photonic quantum technologies described above have relied on large-scale optical elements (such as beam-splitters and mirrors) bolted to room-sized optical tables, with photons propagating through air as shown in Fig.~\ref{Fig_bulkTakeuchi}. Moreover, these approaches do not offer miniaturisation and scalability that photonic QIS will require. On-chip optical waveguides hold great promise: near-perfect mode overlap; no optical interfaces (the major contributor to loss); miniaturisation and hence scalability; and the promise of integration with single-photon sources and detectors, and have begun to be implemented.

Because the scheme proposed by KLM and most of the integrated gates reported in the literature are based on spatial qubit encoding, we start by providing a short introduction to this and then take a look at some relevant examples. In this context, we first focus on the realization of a probabilistic CNOT gate since it is an elementary building block allowing the realization of a general purpose quantum computer. In a second part, we detail a more advanced IO quantum circuit consisting of several one- and two-qubit gates that was recently used to perform a compiled version of Shor's quantum factoring algorithm~\cite{Shor_QComp_1994,AlbertoPoliti09042009}. Finally, we describe a particular experiment of the quantum random walk since it clearly highlights the kind of realization that only IO systems can enable.

\subsection{Spatial encoding to create qubits}

\begin{figure}[h!]
\centering%
\includegraphics[width=\columnwidth]{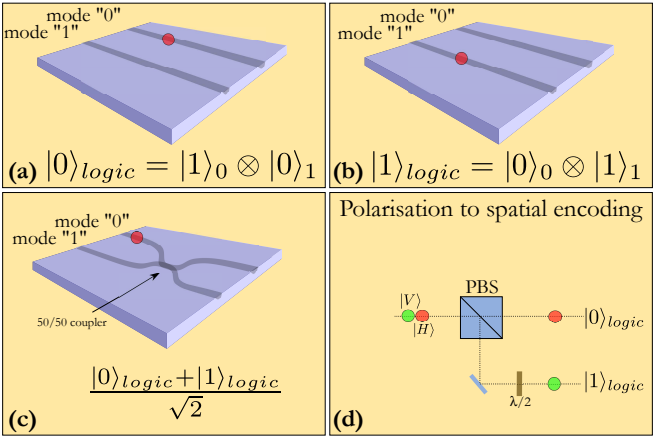}
\caption{Qubit spatial encoding. (a) and (b) There are two channels, labelled $0$ and $1$, enabling the encoding of a qubit on one spatial mode of the chip. (c) A simple beam-splitter allows producing the coherent superposition state $\frac{\ket{0}+\ket{1}}{\sqrt{2}}$. (d) Transformation of a polarization qubit to spatial qubit.\label{Fig_pedago}}
\end{figure}

In linear optical quantum computing, the qubit of choice is usually taken to be a single photon that can populate two different modes coherently: $\ket{0}_{logic}=\ket{1}_0\otimes\ket{0}_1$
and $\ket{1}_{logic}=\ket{0}_0\otimes\ket{1}_1$ where each ket corresponds to a spatial mode and indicates the number of photons occupying the mode. This is called a dual-rail qubit. For instance, when the two modes represent the internal polarization degree of freedom of the photon $\ket{0}_{logic}=\ket{1}_H\otimes\ket{0}_V=\ket{H}$ and $\ket{1}_{logic}=\ket{0}_H\otimes\ket{1}_V=\ket{V}$, we speak of a polarization qubit~\footnote{If the two computational qubit values are ``early'' and ``late'' arrival times in a detector, it is called time-bin encoding.}. In this review we will reserve the term ``dual rail'' for a qubit with two spatial modes. Here the two modes are two channels and it is important to remember that these two representations are mathematically equivalent and we can physically switch between them using polarization beam splitters (see Fig.~\ref{Fig_pedago} for some graphical examples).

IO is particularly suitable for spatial encoding schemes since it allows implementing many channels that can be made to switch, recombine or interfere at specific locations. Other favourable aspects of the IO implementation are:
\begin{itemize}
 \item The compactness and the monolithic construction of the chip ensure negligible phase drift between the channels during propagation;
 \item A spatial Hadamard gate, which is crucial for quantum processing, is a simple 50/50 directional coupler that was realized with IO technology over 30 years ago~\cite{Ostrow_75}. 
\end{itemize}
On the other hand, the natural birefringence of such waveguides usually complicates using polarisation encoding since the decoherence of the qubit is faster than the propagation time along the circuit and would required an integrated correction stage.

\subsection{CNOT gates}

A major difficulty for optical quantum information processing is the realization of two-qubit-entangling logic gates. The canonical example is the controlled-NOT (CNOT) gate, which flips the state of a target qubit only if the control qubit is in the state '1'. This is the quantum analogue of the classical XOR gate. Following Fig.~\ref{Fig_CNOT}, the two optical paths that encode the target qubit are combined at a 50/50 directional coupler, and the output is then combined at a second coupler to form a Mach-Zehnder interferometer. The logical operation of this device leaves the qubit unchanged: $\ket{0}\rightarrow\ket{0}$; $\ket{1}\rightarrow\ket{1}$. If, however, a phase shift is applied inside the interferometer (such that $\ket{0}+\ket{1}\rightarrow \ket{0}-\ket{1}$) the target qubit undergoes a ``bit-flip'', or NOT operation, $\ket{0}\leftrightarrow\ket{1}$. A CNOT gate must therefore implement this phase shift if the control qubit is in the '1' path.

\begin{figure}
\centering%
\includegraphics[width=\columnwidth]{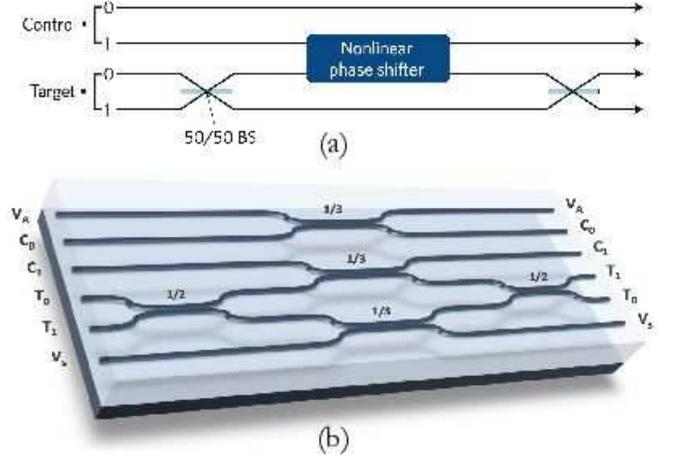}
\caption{(a) Conceptual schematic of a CNOT logic gate, and (b) its realization as an integrated chip based on KLM proposal. The KLM gate is probabilistic (1/9) and $V_A$, $V_B$ are unused ports which are required to ensure that the global success rate remains 1/9. Figures extracted from Refs.~\cite{O'Brien09,Laing10}, respectively.\label{Fig_CNOT}}
\end{figure}

Remarkably, the use of an unbalanced directional coupler ($\eta=1/3$), within which two photons can interfere, leads to the right phase shift and is called a CZ gate, but the price to be paid is a probability of success of 1/9. This is basically the principle of the CNOT gate demonstrated in 2010 by a Bristol team~\cite{Laing10}. When considering the chip of Fig.~\ref{Fig_CNOT}, one has to note the presence of two additional $1/3$-couplers (at the top and bottom) to balance the losses and ensure the unitarity of the gate. The authors used an SOS waveguide circuit. Waveguides are brought within several $\mu$m proximity to realize directional couplers whose reflectivity $\eta$ can be controlled via the length of the coupling region. 

Standard SOS optical lithography fabrication techniques allows obtaining a refractive index contrast of $\delta n= 0.5\%$ to give single-mode operation at 804\,nm for 3.5 by 3.5\,$\mu$m waveguides which are comparable to the core size of conventional single-mode optical fibers at 800\,nm, and allows good coupling of photons to fiber-pigtailed single-photon sources and detectors. The separation of the waveguide in the coupling region was chosen to be $4\,\mu$m. This value, in combination with the values of the refractive indices of the core and cladding and dimensions of the core, give a coupling length of $L_c = 3100$\,$\mu$m. The value of the radius of curvature of bends $R = 15000$\,$\mu$m was chosen to satisfy the requirements of keeping the devices compact and minimize losses. The sample is few centimeter-long and exhibits losses of typically $\sim$0.1\,dB/cm while the overall coupling efficiency was shown to be around 80$\%$ in and out of the device. 

The two-photon interference on such a device when the coupling ratio of the coupler was choosen to be $\eta=0.5$ has shown quantum interference visibility of $0.995\pm0.004$. When switching to the full CNOT circuit, they have measured using quantum tomography the truth table of the device and obtained a fidelity of $0.969\pm0.002$ with ideal output of the CNOT operation.  

Using the same technology, it is possible to build some more complex circuits by interconnecting a few CNOT gates together, as presented in the following section.

\subsection{Shor's Quantum Factoring Algorithm}

If not the most demanding, the experimental demonstration of Shor's algorithm had a large impact. This algorithm was the first that actually suggested the potential of quantum computing for an applied task: factoring the product of two prime numbers exponentially faster than any known conventional method~\cite{Shor_QComp_1994}.
Its full version is designed to factorise any given input. However, for the presented proof-of-principle demonstration~\cite{Politi09} (and all previous ones based on bulk optics~\cite{Lanyon07} or nuclear magnetic resonance~\cite{Vandersypen_ShorAlgo_2001}), it is a compiled version of the algorithm which is used. This means that it only works for a given input: 15. Even if it is a much simpler algorithm that only requires 4 qubit inputs experiencing 2 CZ gates and 6 additional Hadamard gates, it is still a demonstration of a small scale device constituting the embryon of, hopfully, much more complex large-scale devices.

\begin{figure}
\centering%
\includegraphics[width=\columnwidth]{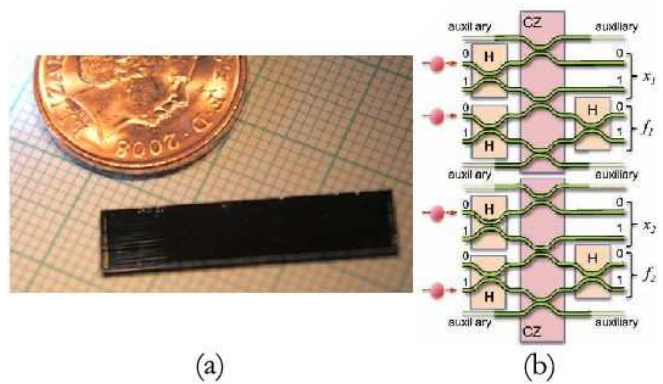}
\caption{Shor's quantum factoring algorithm on a photonic chip. Figures extracted from Ref.~\cite{Politi09}.\label{Fig_Shor}}
\end{figure}

In more detail, the chip, depicted in Fig.~\ref{Fig_Shor}, implements the quantum order-finding routine that enables it to factorize $N=pq$. This quantum routine finds the order of $a\,\mbox{mod}\,pq$, defined to be the smallest integer $r$ that satisfies $a^r\equiv 1\,\mbox{mod}\,pq$. It is then straightforward to find the prime factors from the order with a classical computer. For the case discussed here, the authors choose $a=2$, the quantum routine finds $r=4$, and the prime factors $p$ and $q$ are given by the greatest divisor common to $a^\frac{r}{2} \pm 1$ and $N$, which leads to 3 and 5, respectively. This is the circuit demonstrated in Ref.~\cite{Politi09}.

Such a device needs two photon-pair sources at its input and produces a valid  output (99$\pm$1\% fidelity) whenever a fourfold coincidence occurs between the gates $x_1$, $f_1$, $x_2$, and $f_2$. In the experiment the coincidence rate was about 100\,Hz and the desired four-fold coincidences occurred with the expected 1/9 probability.
The high quality of the results shows the successful implementation of a small-scale quantum algorithm. It demonstrates the feasibility of executing complex, multiple-gate quantum circuits involving coherent multiqubit superpositions of data registers.

To further demonstrate the potential of integrated photonics for quantum processing, we now present an experiment that was only feasible using IO waveguides.

\subsection{Quantum random walk circuit}

It has been theoretically proven that Quantum Random Walks (QRW)~\cite{Aharonov93} allow the speed-up of search algorithms~\cite{Shenvi03,Potocek09} and the realization of universal quantum computation~\cite{Childs09}. A QRW can be implemented via a constant splitting of photons into several possible sites. Basically, chaining 50/50 beam-splitters is the best way to simulate a QRW but the price to be paid is the quickly growing size of the set-up. Considering that the major challenge is the need for a low-decoherence system that preserves the nonclassical features of the photons, up to now the largest realization using bulk optics was limited to a 5 layer system~\cite{Do05,Schreiber10}.

\begin{figure}
\centering%
\includegraphics[width=\columnwidth]{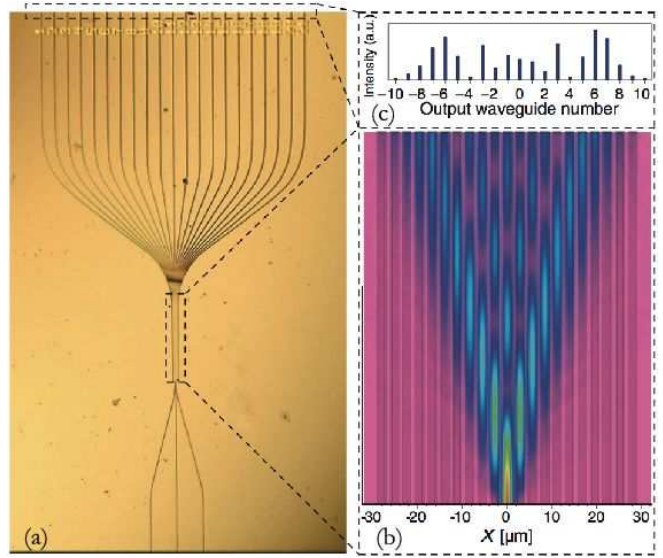}
\caption{Quantum Random Walk on a photonic chip. Figures extracted from Ref.~\cite{Peruzzo10}.\label{Fig_QRW}}
\end{figure}

The basic idea could be to use the same SOS structures as described previously for the Shor algorithm. However, the low refractive index contrast ($\delta = \frac{n_{core}^2 - n_{cladding}^2}{2n_{core}^2} \approx 0.5\%$) in this architecture results in a large minimum bend radius ($\leq0.1\,dB$ loss at 800\,nm) of $\sim$15\,mm, making it unsuitable for coupling into and out of large-array quantum walk devices. A solution is to use $SiO_xN_y$ (silicon oxynitride), a material that enables a much higher refractive index contrast than SOS, resulting in more compact devices and a practical means to realize large coupled waveguide arrays that can be coupled to optical fibers. The device shown in Fig.~\ref{Fig_QRW} is a 5-mm-long silicon chip with $SiO_xN_y$ waveguides with high refractive index contrast $\Delta= 4.4\%$. The minimum bend radius for this index contrast is 600\,$\mu$m, which enables much more rapid spreading of the waveguides before and after the evanescent coupling region. In the coupling region the waveguides are separated by 2.8\,$\mu$m, whereas at the input and output of the chip they are separated by 250 and 125\,$\mu$m for photon injection and collection, respectively, imposed by standard fiber arrays.

The device, demonstrated in 2010 by the Bristol team, merges, on a single chip, 21 continuously evanescently coupled waveguides (represented on the right side of Fig.~\ref{Fig_QRW}) which allows advancing the physics of the implementation of a QRW. Using such a chip, the author were able to test classical random walks using single photons and QRWs using photon-pairs prepared in various entangled states in order to fully explore the physics underlying QRW. From the measurement point of view, they compare theoretical output results with the experimental ones and have shown an agreement of more than 0.90.

\begin{figure}
\centering%
\includegraphics[width=\columnwidth]{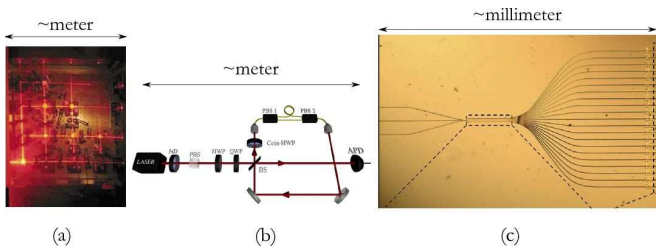}
\caption{Size and complexity comparison of (a) bulk optics~\cite{Do05}, (b) semi-fibred~\cite{Schreiber10} and (c) integrated~\cite{Peruzzo10} QRW experiments. It also has to be noticed that setup (a) and (b) only implement 5 layers of random walk, whereas setup (c) managed to integrate 21 layers of random walk. Figures extracted from Refs.~\cite{Do05,Schreiber10,Peruzzo10}, respectively.\label{Fig_compa}}
\end{figure}

Note the reduction of size that the IO random walk scheme allows compared to bulk and semi-fibred solutions, respectively, as depicted in Fig.~\ref{Fig_compa}.

\subsection{3D circuits and further directions}

\begin{figure}
\centering%
\includegraphics[width=\columnwidth]{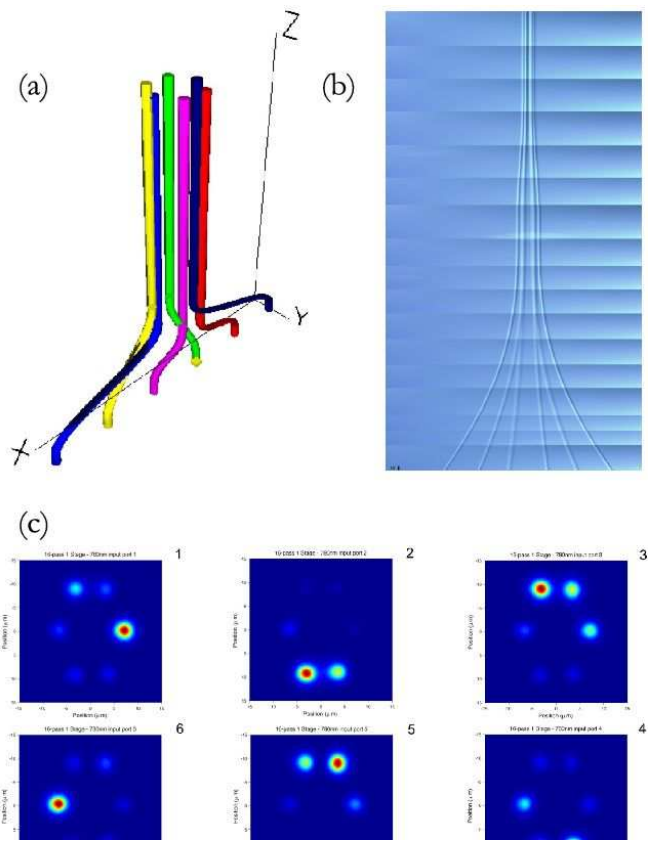}
\caption{Quantum Random Walk on a 3D photonic chip. (a) Image of a six-waveguide tube with a single stage fan-in section. (b) Picture of the waveguides. 
(c) Beam profiles of the six waveguide when injecting light in one of each. Figures extracted from Ref.~\cite{Twamley10}. \label{Fig_3D}}
\end{figure}

3D waveguide technology allows the realization of previously unfeasible waveguide circuits such as quantum random walks on a ring. In 2010, an Australian group from Macquarie has created an integrated photonic waveguide network~\cite{Twamley10} using the laser direct-write fabrication technique. This architecture exploits the unique three-dimensional advantage which the direct-write fabrication technique has over the lithographic approach, albeit at the price of linear, as opposed to parallel, processing of the chip. Six parallel waveguides are created in a tubular array, as shown in Fig.~\ref{Fig_3D}. In terms of comparison, such a structure has a coupling network equivalent to approximately 96 bulk optics beam splitters but are compressed to span a physical distance of merely 20-22\,mm.

Finally, we conclude this aspect of the review with a recent proposal towards achieving deterministic quantum processing~\cite{Saleh:10}. The scheme proposes mixing spatial and polarization encoding using Ti-indiffused channel $LiNbO_3$ waveguides. Contrary to previously discussed chips with spatial qubits encoded over several single mode waveguides, the authors theoretically investigate the use of modal and polarization qubits within a two-mode waveguide. They show that basic transformations of modal qubits are possible such as mode rotators and the modal Pauli spin operator $\sigma_z$. Based on those tools, they describe the design of a \textit{deterministic}, two-qubit, single-photon CNOT gate.

\section{Perspectives and conclusions}

One of the major directions for future development is clearly increased functional integration. Such integration could help resolve the continued problem of input/output coupling losses but certainly not solve it completely. Hopefully, as integrated quantum optical components begin to enter the real world, a much larger effort will be applied to this technical issue. There is no fundamental reason preventing a solution to this problem.

\begin{figure}
\centering%
\begin{tabular}{c}
\includegraphics[width=\columnwidth]{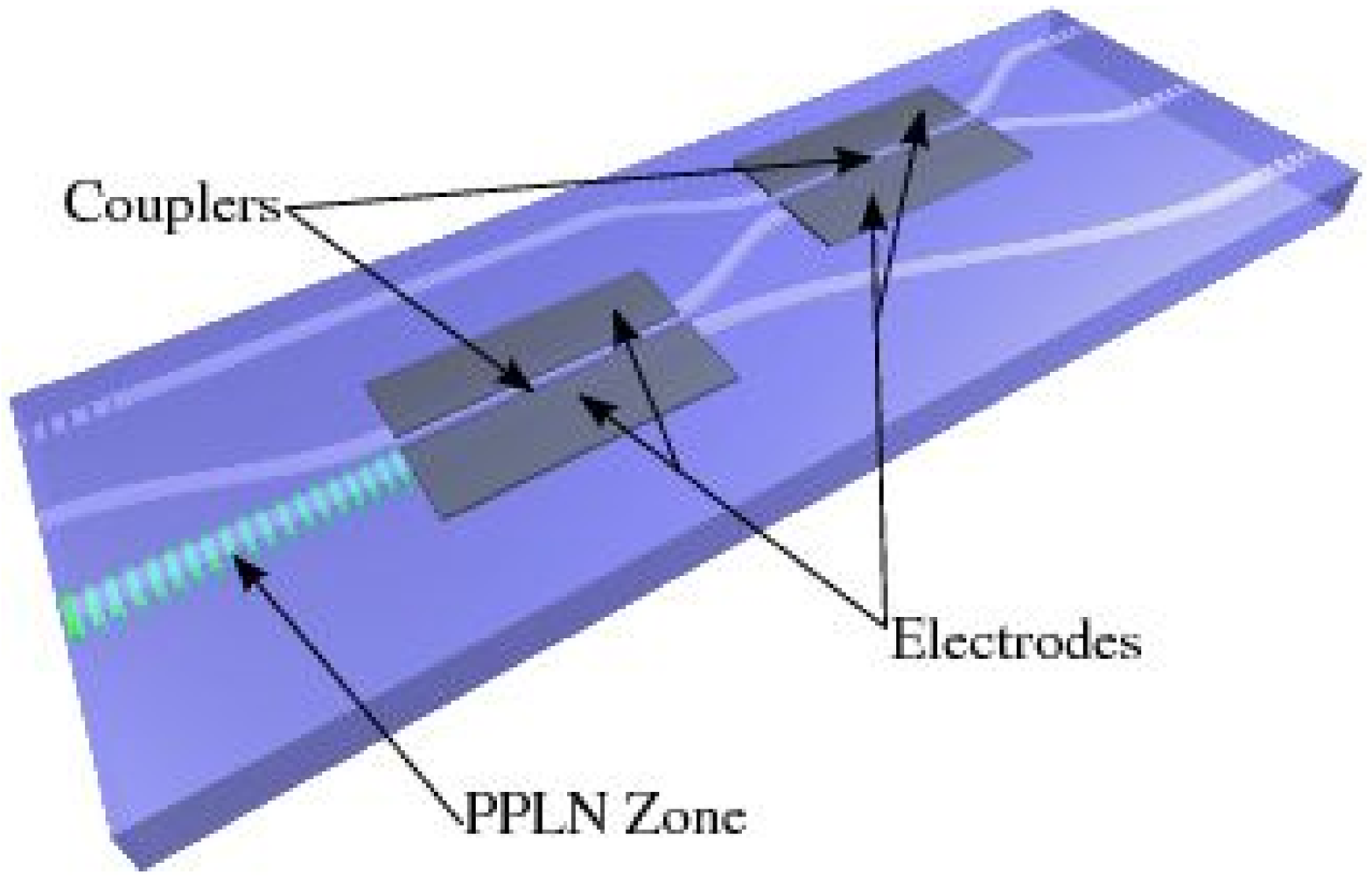}\\
(a)\\ ~\\
\includegraphics[width=\columnwidth]{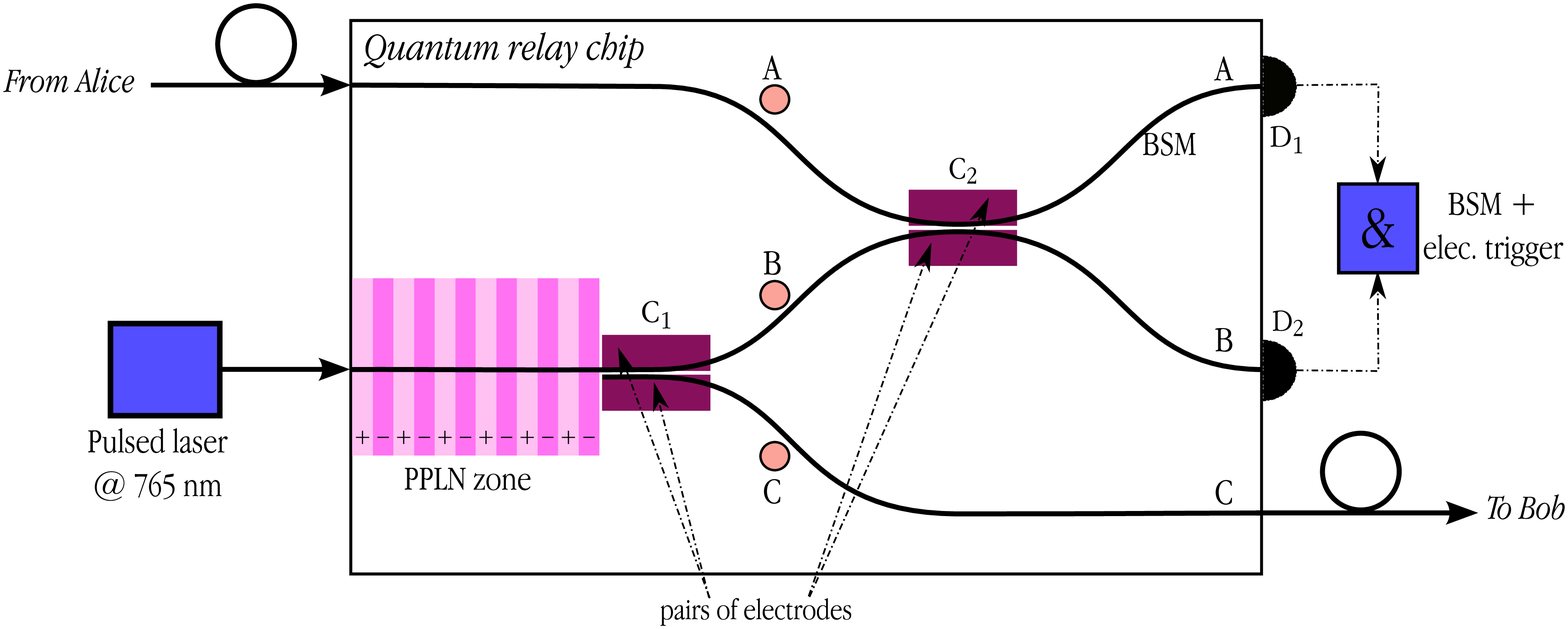}\\
(b)
\end{tabular}
\caption{\label{Fig_QrelayChip_Nice}(a) 3-D representation of a quantum relay chip made for extending one-way quantum cryptography systems. (b) Schematics of the quantum relay chip with its two electro-optically controllable couplers $C_1$ and $C_2$. $D_1$ and $D_2$ are two detectors responsible for the Bell state measurement (BSM). At the end of the quantum channel, Bob's detector is triggered by the AND-gate (\&) placed after these two detectors.}
\end{figure}

As an example of higher integration, we will discuss a quantum relay chip applicable to long-distance quantum communication. In this case, IO on lithium niobate permits realizing a telecom-like quantum relay chip that could provide the relay function, in a compact, stable, efficient, and user-friendly fashion.
Figure \ref{Fig_QrelayChip_Nice} presents two schematics of a chip designed and fabricated in our group on which all the necessary optical functions are merged for implementing teleportation on-a-chip. We will now discuss how this chip works.

Let's suppose there's an unknown qubit $1$, encoded on a photon at 1535\,nm and travelling along a fiber quantum channel connected at port $A$ to the relay chip. At the same time, a photon from a laser pulse at 765\,nm, synchronized with the arrival time of qubit $1$, enters through the non-linear zone of the chip which consists of a waveguide integrated on PPLN. Thanks to the appropriate choice of the periodic poling grating period (here around 16\,$\mu$m), this pump photon can be converted by SPDC into an entangled pair of photons ($2$ and $3$) whose wavelengths are also centred at 1535\,nm.
Then, the first 50/50 directional coupler ($C_1$) is used to separate the created entangled photons in such a way that photons $1$ (sent by Alice) and $2$ arrive at the same time at the second 50/50 coupler ($C_2$). If conditions on the polarization states, central wavelengths, and coherence times are met, photons $1$ and $2$ can be projected onto one of the four entangled ``Bell states'' (BSM). The resulting state is identified by detectors $D_1$ and $D_2$ placed at the output of the chip. As a result of this measurement, the qubit initially carried by photon $1$ can be teleported to photon $3$ that exits the chip at port $B$, in theory without any loss of its quantum properties. This is made possible since photon $3$ was initially entangled with photon $2$. As a consequence the resulting electrical trigger is not only the signature of the presence of the initial carrier at the relay chip location but also of the departure of a new one encoded with the same qubit state that remains unknown.

From a practical point of view, when this teleportation process is repeated on all the qubits travelling along the quantum channel, we obtain, at the output of the chip, qubits again encoded on photons at 1535\,nm. But now, each arrives synchronized with an electrical signal given by the BSM at detectors $D_1$ and $D_2$ which allows triggering Bob's detectors at the end of the channel and therefore increasing both the SNR and the communication distance.

On the technological side (see Fig.~\ref{Fig_QrelayChip_Nice}), the chip features a photon-pair creation zone, two $50/50$ couplers, and segmented tapered waveguides~\cite{Castaldini_09} at all input/output ports to maximize the mode overlap between the fibers and the waveguides at the particular wavelength of 1535\,nm.
The waveguiding structures, obtained using soft proton exchange, enables light propagation with low losses ($\simeq$0.3\,dB/cm) and a very high conversion efficiency in the PPLN zone thanks to strong light confinement over a cm-long distance ($\delta n \simeq 2.2\cdot10^{-2}$).
The directional couplers $C_1$ and $C_2$ consist of two waveguides integrated close to each other over mm-long distances. If the spacing is sufficiently small, energy is exchanged between the two guides. An electro-optical control of the coupling ratio is made using deposited electrodes enabling switching from 0 to 100\% coupling ratios and reconfigurable operation. To correctly separate photons $2$ and $3$ and entangle $1$ and $2$, the two couplers are set to the 50/50 ratio.

We proceeded to both classical and quantum characterizations of the chip in the single photon counting regime. We could verify that the PPLN zone emits paired photons around the 1535\,nm desired wavelength, that the two couplers can actually be set at the 50/50 ratio, and that the input/output insertion losses, including in and out coupling, straight and bending propagation losses are about 7\,dB. Quantum experiments using a single photon input at the upper-left input port and pumping synchronously with a mode-locked laser at 765\,nm are underway.

Others majors directions include the integration of cavities and photonic crystal structures with quantum dots~\cite{dousse_Qdotultrabright_2010,Rivoire_dot_11}, NV centers in diamond~\cite{Prawer_Diamond_review_2008}, and the development of silicon nano-wire quantum devices~\cite{sharping_generation_2006,harada_generation_2008,Clemmen_Gene_Nano_2010}.

In conclusion, the impact of Integrated Optics on modern quantum optics has been considerable. Demonstration of IO applications to quantum sources, detectors, interfaces, relays, memories, repeaters and linear optical quantum computing have been described in this review. By their inherent compactness, efficiencies, and interconnectability, many of the demonstrated individual devices can clearly serve as building blocks for more complex quantum systems, that could also profit from the incorporation of other guided wave technologies. Elementary implementations of on-chip integration has begun and should be greatly expanded in the future. While the input and output coupling continues to remain a difficult problem for integrated optics, the continuing evolution of these developments, as well as the incorporation of recently presented new technologies, such as nano-wire waveguides and photonic crystal circuits, should add considerable impetus to this rapidly evolving field.

\section*{acknowledgement}
We deeply thank K. Thyagarajan and W. Sohler for fruitful discussions. We acknowledge the European ERA-SPOT program ``WASPS'', the European ICT-2009.8.0 FET Open program for the ``QUANTIP'' project (grant agreement 244026), the Agence Nationale de la Recherche (ANR) for the ``e-QUANET'' project (grant ANR-09-BLAN-0333-01), the CNRS, the Universit\'e de Nice -- Sophia Antipolis, the Conseil Regional PACA, and the Minist\`ere de l'Enseignement Sup\'erieur et de la Recherche for financial support.

\end{document}